\documentclass[aps,prl,reprint,superscriptaddress,twocolumn,showkeys,amsmath,amssymb,longbibliography,floatfix]{revtex4-2}
\usepackage{amsmath,amssymb,bbm,mathrsfs,bm,braket,color,graphicx,comment,amsfonts,dsfont}
\usepackage[colorlinks,linkcolor=blue,citecolor=blue,urlcolor=blue]{hyperref}
\usepackage[mathscr]{euscript}
\usepackage{physics}
\usepackage{xcolor}
\usepackage[normalem]{ulem}
\usepackage{bm}
\usepackage{orcidlink}
\usepackage{multirow}
\usepackage{microtype}
\usepackage[capitalise]{cleveref}
\usepackage{times}

\usepackage{pifont}

\newcommand{\sxx}{\sigma_{xx}}
\newcommand{\sxy}{\sigma_{xy}}

\newcommand{\chitwoD}{\chi^{\mathrm{2D}}}
\newcommand{\rd}{\mathrm{d}}
\newcommand{\ii}{\mathrm{i}}
\newcommand{\Eg}{E_{g}}
\newcommand{\half}{\tfrac{1}{2}}

\begin{document}

\title{Quantum-geometric bounds on Casimir repulsion}

\author{Adolfo G. Grushin\orcidlink{0000-0001-7678-7100}}
\email{grushin@dipc.org}
\affiliation{Donostia International Physics Center, P. Manuel de Lardizabal 4, 20018 Donostia-San Sebastian, Spain}
\affiliation{IKERBASQUE, Basque Foundation for Science, Maria Diaz de Haro 3, 48013 Bilbao, Spain}
\date{\today}

\begin{abstract}
The quantum geometric tensor has been shown to bound the gap, optical absorption, and dielectric susceptibilities of materials. Here we derive new quantum-geometric bounds on the magnitude and sign of the Casimir force between two-dimensional plates in the long-distance limit. 
These bounds limit the previously attributed benefit of increasing the plate's Chern number to maximize repulsion, and give a quantum geometric origin to the stronger attractive force of metallic plates, regardless of their Chern number. These bounds allow us to infer that flat Chern bands that saturate geometric bounds, including Landau levels and moiré flat bands, enlarge the window where Casimir repulsion exists and bring the repulsive crossover to smaller, more experimentally relevant distances. We derive estimates for material platforms such as twisted MoTe$_2$. 
Our work shows that quantum-geometric bounds constrain repulsive Casimir forces beyond previously known theorems, and suggests new optimization strategies to observe repulsion.
\end{abstract}

\maketitle

\textit{Introduction.-} 
A convenient way to organize information about a material's wavefunctions is using the quantum geometric tensor~\cite{provost80,berry84}. Its real part, the quantum metric, captures the wavepacket's spread, and its imaginary part, the Berry curvature, captures the wavepacket's anomalous velocity, which determine properties of modern quantum materials~\cite{yu-npjquantummater25,verma2025quantumgeometryrevisitingelectronic,Gao2025quantumgeometry}. The quantum geometric tensor elegantly captures the existence of sum-rules and bounds on physical observables~\cite{souza-prb00,aebischer-prl01,martin-book04,roy-prb14,peotta-natcomms15,ozawa-prb21,onishi-prx24,komissarov-natcomms24,onishi-prb24,verma-pnas25,souza-scipost25,onishi-prr25,Passos2026}, like the dielectric susceptibility or the electrical conductivity. 

The electrical conductivity determines the ability of a body to reflect electromagnetic waves through the reflection matrices~\cite{DG02}.
These matrices determine the boundary conditions on the electromagnetic field between two bodies, whose energy then depends on the distance $d$ between them. Such energy-distance dependence results in a force, the Casimir force~\cite{Cas48,Lifshitz1956,DLP61,LLv8,BKMM09,Milton_2012,Woods2016}, which can be either attractive or repulsive. 

Finding situations where the Casimir force is repulsive rather than attractive is a long-standing problem with practical and fundamental implications~\cite{Milton_2012,Woods2016}.
Practically, repulsion can help overcome undesired sticking of nano-mechanical components in devices~\cite{Mun09b,IKJ13}.
Fundamentally, finding a repulsive Casimir force is challenging because of the restrictive theorems that force attraction in vacuum:
the force is attractive when the bodies are related by reflection~\cite{Kenneth2006,B06},
and a stable equilibrium is ruled out for dielectric bodies~\cite{Lam97,RK09,RE09}.
The literature is rich in proposals that aim to evade these theorems by using intermediate dielectrics~\cite{DLP61,Mun09b,Zwo10},
geometries~\cite{LMR10}, surface plasmons~\cite{Lam08,Pir09},
magnetic or magneto-electric~\cite{B74,Metalidis2002,Metalidis2010,Rosa2010,Rosa08} materials,
meta-materials~\cite{Zhao2009,Zhao2010,Deng2011},
and topological phases of matter~\cite{BV00,GC11,GRC11,Rodriguez-Lopez2011,Chen2011,TseMacDonald2012,Grushin2012,Rodriguez-Lopez2014,Wilson2015,RLKortKamp2017,FKV2018,Farias2020,Vassilevich2020}.

\begin{figure}[t]
  \includegraphics[width=\columnwidth]{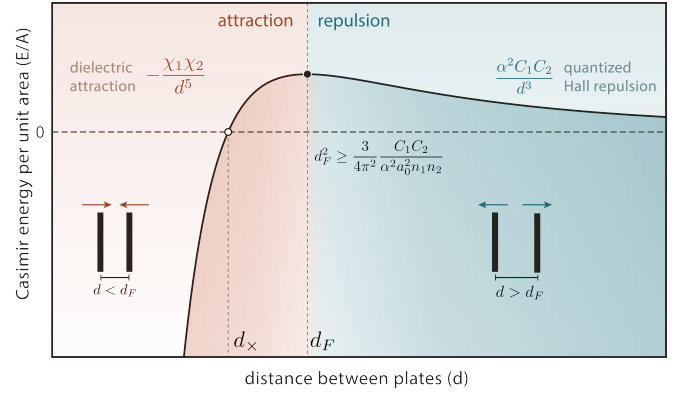}
  \caption{Schematic Casimir energy per area $E/A$ between two
  Chern-insulating sheets with Chern numbers $C_1C_2>0$ and densities $n_{i=1,2}$ as a function of separation $d$. At short
  distances the dielectric response dominates and the interaction is
  attractive; beyond the force-crossover distance $d_F$, where $E/A$
  is maximal, the Hall response drives Casimir repulsion. Quantum
  geometry bounds $d_F$ from below by Eq.~\eqref{eq:main-bound}. Within the
  leading large-distance description, repulsion is excluded at separations
  smaller than this bound.
  }
  \label{fig:schematic}
\end{figure}

While these examples showcase that repulsion is possible on a case-by-case basis, it is desirable to establish system-agnostic conditions that can lead to repulsion. Ref.~\cite{Rosa2010} established an important step by deriving the conditions that the penetration depths associated with each plate should satisfy to guarantee repulsion. The benefit of these bounds is that they do not assume a specific system, but the drawback is that the numerical prefactors that determine them are not easily interpreted.

Here we use the bounds on observables related to the quantum geometric tensor to derive novel bounds on the sign and magnitude of the Casimir force. We focus on two-dimensional insulating plates with gap $E_g$ and a DC Hall conductivity $\sigma_{xy}(0)=C\frac{e^2}{h}$. The result is neatly summarized by the following bound 
\begin{equation}
d^2_F\ge\frac{3}{4\pi^2\alpha^2}\,
\frac{C_1 C_2}{(a^{*}_{\perp})^{2}\,n_1 n_2}, \qquad C_1C_2>0.
\label{eq:main-bound}
\end{equation}
Here $d_{F}$ is the crossover distance where the attractive Casimir force at small distance necessarily becomes repulsive. The bound relates this distance to the Chern numbers of both plates $C_{i=1,2}\neq 0 $, which should satisfy $C_1C_2>0$ for repulsion to exist. It is expressed in terms of the electronic densities of each plate, $n_{i}$, the fine structure constant $\alpha=1/137$ and the effective Bohr radius $a^{*}_{\perp}=\hbar^{2}/(e^{2}\langle m_{*}\rangle_{\perp})$, set by the in-plane optical mass of the relevant low-energy bands, which reduces to the bare Bohr radius $a_0$ for parabolic bands with the free-electron mass $m_e$. The bound Eq.~\eqref{eq:main-bound} is a consequence of a recently found bound on the dielectric susceptibility \cite{Passos2026}. We here show that this result also bounds the attractive part of the Casimir energy from above by the square of the Chern numbers.

Eq.~\eqref{eq:main-bound} should be read as a bound on the onset of repulsion, built from zero-frequency (DC) geometric data alone. Within the leading large-distance truncation it is rigorous: the dielectric attraction can only be stronger than the geometric floor set by the susceptibility bound, so repulsion cannot set in at separations smaller than Eq.~\eqref{eq:main-bound}. 

\begin{figure}[t]
\includegraphics[width=\columnwidth]{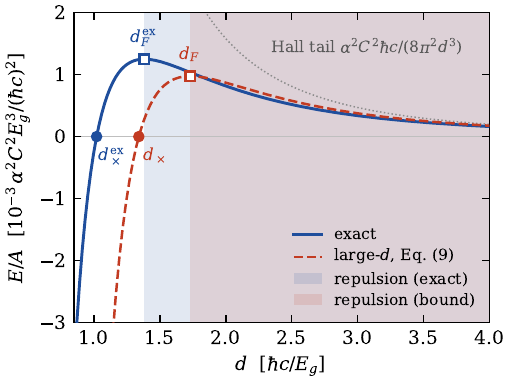}
\caption{Geometric bounds overestimate attraction.
Energy per unit area of two identical Landau-level sheets, with Chern number $C$ and cyclotron gap 
 $\Eg=\hbar \omega_c$, as a function of separation $d$. 
The dashed red curve shows the large-distance Casimir energy, Eq.~\eqref{eq:E-combined}.
The solid blue curve shows the exact result using Landau level optical conductivities. 
Filled circles mark the zeros of the energy,
${d^{\mathrm{ex}}_\times}=1.021\,\hbar c/\Eg$ (exact) and ${d_\times}=3/\sqrt{5}\simeq1.342\,\hbar
c/\Eg$ (large distance); open squares mark the zeros of the force  $F/A=-\partial_d(E/A)$,
${d^{\mathrm{ex}}_F}=1.383\,\hbar c/\Eg$ and ${d_F}=\sqrt{3}\simeq1.732\,\hbar c/\Eg$. Each maximum is an unstable equilibrium separating
the repulsive large-$d$ regime from the attractive short-distance regime.
At large $d$ both curves approach the quantized Hall
tail $\alpha^{2}C^{2}\hbar c/(8\pi^{2}d^{3})$ (dotted). The difference between the shaded areas {shows} that the geometric bound {overestimates} the attractive region.
}
\label{fig:LLcrossover}
\end{figure}

\begin{figure}[t]
  \centering
  \includegraphics[width=0.48\textwidth]{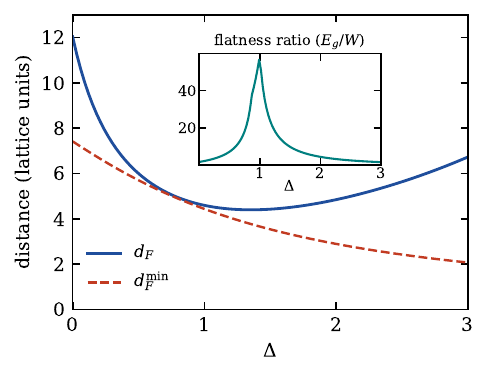}
  \caption{Flat-band-induced increase of the repulsive region.
  We plot the force sign-change distance for two identical $C=3$ flat-band Chern-insulator
  sheets in the three-band model of Refs.~\cite{Passos2026,YGS12}. It is plotted as a function of the model parameter $\Delta$, that controls the flatness ratio of the lowest band with $\Delta=1$ achieving the maximal flatness.  The solid
  line is the value obtained from the Kubo polarizability, Eq.~\eqref{eq:flatband-main-dF};
  the dashed line is the minimum value obtained by inserting the Chern
  susceptibility bound.  Distances are in lattice-constant units $a=1$. We see that when flattening the band helps to saturate the geometric bounds it reduces $d_F$ and increases the region where repulsion can be observed. Inset: flatness ratio, defined as gap $E_g$ over bandwidth $W$, of the lower band as a function of $\Delta$.}
  \label{fig:flatband-main-distance}
\end{figure}

In what follows we derive the bound and illustrate two effects that displace the actual crossover when going beyond the long-distance approximation. First, the full finite-frequency response weakens the attraction relative to its static extrapolation, shortening the crossover distances, thus pushing it to more experimentally accessible distances. Eq.~\eqref{eq:main-bound} is in this sense a conservative bound on the onset of repulsion, an effect we will illustrate in Fig. \ref{fig:LLcrossover}. Second, if the susceptibility bound is not saturated, the crossover is pushed to larger separation. Hence, saturating geometric bounds is beneficial to bring Casimir repulsion to shorter, more observable distances. We illustrate this effect in Fig. \ref{fig:flatband-main-distance}.

\textit{Derivation of the bounds.-} 
The Casimir energy per unit area between two sheets at distance $d$ at zero temperature is given by the Lifshitz scattering formula~\cite{Cas48,Lifshitz1956,DLP61,LLv8,BKMM09}:
\begin{equation}
\frac{E}{A}=\frac{\hbar}{2\pi}\int_0^{\infty}\!\rd\xi
\int\!\frac{\rd^2 q}{(2\pi)^2}\,
\ln\det\!\Big[\mathbbm{1}-R'_1(\ii\xi,q)\,R_2(\ii\xi,q)\,e^{-2\kappa d}\Big],
\label{eq:lifshitz}
\end{equation}
in terms of $\kappa=\sqrt{q^2+\xi^2/c^2}$ and the reflection matrices
\begin{equation}
R_a(\ii\xi,q)=
\begin{pmatrix}
r^{(a)}_{xx}(\ii\xi,q) & r^{(a)}_{xy}(\ii\xi,q)\\[2pt]
r^{(a)}_{yx}(\ii\xi,q) & r^{(a)}_{yy}(\ii\xi,q)
\end{pmatrix}
\label{eq:Rmatrix}
\end{equation}
of each plate $(a=1,2)$, that depend on the imaginary frequency $\omega=i\xi$ and the momentum $q$. The prime in $R'$ accounts for reverting the sign of the scattering direction.
The reflection matrices depend on the conductivity tensor, $\sigma^{(a)}_{ij}$, of each sheet $a$,
\begin{equation}
\sigma^{(a)}_{ij}(\omega)=\sxx^{(a)}(\omega)\,\delta_{ij}+\sxy^{(a)}(\omega)\,\epsilon_{ij},
\label{eq:sigma-tensor}
\end{equation}
where we neglect their momentum dependence ($q\to 0$).
For a gapped two-dimensional insulator, at imaginary frequencies $\omega=i\xi$ much
smaller than the optical gap $\Eg/\hbar$, the Hall conductivity approaches its
quantized dc value and the longitudinal conductivity vanishes linearly in $\xi$:
\begin{align}
\sxy(\ii\xi)&=C_a\,\frac{e^2}{h}+\mathcal{O}\!\left(\frac{\hbar^2\xi^2}{\Eg^2}\right),
\label{eq:sxy-lowfreq}\\
\sxx(\ii\xi)&=\frac{\xi}{4\pi}\,\chitwoD_a(0)+\mathcal{O}(\xi^3).
\label{eq:sxx-lowfreq}
\end{align}
Here $\chitwoD_a(0)$ is the static sheet electric polarizability, which in two
dimensions has the dimensions of length. 
Notably the diagonal and off-diagonal reflection coefficients depend on the diagonal and off-diagonal conductivities. An explicit expression for a two-dimensional plate was derived in Ref.~\cite{TseMacDonald2012}, and is given in Appendix \ref{app:derivation}.

At large $d$, when $
\frac{\chitwoD(0)}{d}\ll1$ and  $d\gg\frac{\hbar c}{\Eg},$ the energy integral is dominated by low frequencies, where the
sheet conductivities are small (see Appendix \ref{app:derivation}). In this limit, we can take two approximations. First, we can truncate each reflection coefficient to linear order in the sheet's conductivity. Second, we can expand the logarithm in powers of $d$
\begin{equation}
\ln\det(\mathbbm{1}-M)\simeq-\mathrm{Tr}\,M
=-\mathrm{Tr}\!\left[R_1 R_2\right]e^{-2\kappa d}.
\label{eq:leading-log}
\end{equation}
Inserting \eqref{eq:leading-log} into the Lifshitz formula \eqref{eq:lifshitz} gives the leading contribution
\begin{equation}
\frac{E}{A}=-\frac{\hbar}{2\pi}\int_0^{\infty}\!\rd\xi
\int\!\frac{\rd^2 q}{(2\pi)^2}\,e^{-2\kappa d}\,\mathrm{Tr}\!\left[R_1 R_2\right] + \cdots.
\label{eq:E2}
\end{equation}
Performing the phase space integrals (see Appendix \ref{app:derivation}) we get
\begin{equation}
\frac{E}{A}=
-\frac{9}{160\pi^2}\,\frac{\hbar c}{d^5}\,\chi_1\chi_2
+\frac{\hbar c\,\alpha^2}{8\pi^2 d^3}\,C_1 C_2+\cdots,
\label{eq:E-combined}
\end{equation}
Since the Casimir pressure is $F/A=-\partial_d (E/A)$, the first term is attractive, and is the only contribution when $C_{1,2}=0$. Its coefficient coincides with reported long-distance results, see e.g., the appendix of Ref.~\cite{TseMacDonald2012}. The second term is of opposite sign and hence repulsive when $C_1C_2>0$. It has also been derived before~\cite{TseMacDonald2012,Rodriguez-Lopez2014,FKV2018} and dominates in the large $d$ limit~ 
\footnote{An earlier calculation of the Casimir effect between Chern insulators
\cite{Rodriguez-Lopez2014} reported the opposite sign convention, repulsion for \emph{opposite}-sign
Chern numbers. As pointed out in Ref.~\cite{FKV2018} (Sec.~III\,B), that result used
the Lifshitz kernel with $R$ in place of $R'$ in the TE/TM basis; for a Hall plate
this is equivalent to reversing the sign of one Hall conductivity, and after the
correction the result agrees with Refs.~\cite{TseMacDonald2012,FKV2018} and with the
present work. We adopt the corrected sign throughout.}. 

Our goal is to apply geometric bounds to Eq.~\eqref{eq:E-combined}. 
The relevant bound that constrains the first term was derived by Passos and Souza and reads~\cite{Passos2026}
\begin{equation}
\chitwoD(0)\ge\frac{C^2}{\pi a_0 n_e}\equiv\chi_{\min}(C,n_e).
\label{eq:susc-bound}
\end{equation}
This is Eq.~(109) of Ref.~\cite{Passos2026}, where the Bohr radius $a_0=\hbar^2/(m_e e^2)$. For a tight-binding or low-energy
model the bare Bohr radius is replaced by a renormalized value
$a^*_\perp=\hbar^2/(e^2\langle m_*\rangle_\perp)$ set by the in-plane effective optical mass~\cite{Passos2026}. From here on we write the bounds in terms of $a^{*}_{\perp}$, which reduces to $a_0$ for parabolic bands with the bare electron mass.

Because the dielectric energy is proportional to $-\chi_1\chi_2$
with $\chi_a>0$, the bound~\eqref{eq:susc-bound} translates directly into a lower
bound on the \emph{magnitude} of the attractive dielectric term
\begin{equation}
\frac{E_{\mathrm{diel}}}{A}\le
-\frac{9}{160\pi^4}\,\frac{\hbar c}{d^5}\,
\frac{C_1^2 C_2^2}{{(a^{*}_{\perp})^{2}}\,n_1 n_2}.
\label{eq:Ediel-bound-signed}
\end{equation}
Adding  to this bound the remaining term of Eq.~\eqref{eq:E-combined}, the Hall contribution, the constrained large-distance energy satisfies
\begin{equation}
\frac{E}{A}\lesssim
-\frac{9}{160\pi^4}\,\frac{\hbar c}{d^5}\,\frac{C_1^2 C_2^2}{{(a^{*}_{\perp})^{2}} n_1 n_2}
+\frac{\hbar c\,\alpha^2}{8\pi^2 d^3}\,C_1 C_2 .
\label{eq:E-constrained}
\end{equation}
A crossover distance $d_\times$ is obtained by equating the two contributions in
\eqref{eq:E-combined}.
Using $\chi_a\ge\chi_{\min}^{(a)}=C_a^2/(\pi {a^{*}_{\perp}} n_a)$ bounds the crossover distance from below:
\begin{equation}
d_\times^2\ge\frac{9}{20\pi^2\alpha^2}\,
\frac{C_1 C_2}{{(a^{*}_{\perp})^{2}}\,n_1 n_2}.
\label{eq:dcross-bound}
\end{equation}
For identical sheets ($C_1=C_2\equiv C$, $n_1=n_2\equiv n$),
\begin{equation}
d_\times\ge\frac{3}{\sqrt{20}\,\pi}\,\frac{|C|}{\alpha\,{a^{*}_{\perp}}\,n}.
\label{eq:dcross-identical}
\end{equation}
The crossover $d_\times$ locates the zero of the Casimir energy. The experimentally measured pressure, $F/A=-\partial_d (E/A)$, changes sign at a larger separation: 
\begin{equation}
d_F=\sqrt{\tfrac{5}{3}}\,d_\times\simeq1.29\,d_\times,
\qquad
d_F\ge\frac{\sqrt{3}}{2\pi}\,\frac{|C|}{\alpha\,{a^{*}_{\perp}}\,n}
\label{eq:dF-bound}
\end{equation}
for identical sheets. For two different sheets, the bound is that of Eq.~\eqref{eq:main-bound}, and is the main result of this work. 
At $d_F$ the force vanishes, separating repulsion and attraction at larger and shorter distances, respectively. The distance $d_F$ marks an unstable equilibrium, a local maximum of $E(d)$. 
The geometric bound guarantees that a repulsive pressure, if exists, appears at a separation larger than $d_F=\sqrt{5/3}\,d_\times$.

From this expression we can already read that increasing the Chern number can be counterproductive for observing repulsion: it pushes the repulsive window out to larger, less experimentally accessible separations, rather than enhancing the repulsion.

The same truncation bounds not only where repulsion can set in, but also how strong it can be. Repeating for the pressure the steps leading to Eq.~\eqref{eq:E-constrained},
\begin{equation}
\frac{F}{A}\le
-\frac{9}{32\pi^{4}}\,\frac{\hbar c}{d^{6}}\,
\frac{C_1^{2}C_2^{2}}{(a^{*}_{\perp})^{2}\,n_1 n_2}
+\frac{3\hbar c\,\alpha^{2}}{8\pi^{2}d^{4}}\,C_1 C_2 .
\label{eq:F-constrained}
\end{equation}
The right-hand side is maximal at $d_{\mathrm{pk}}=\sqrt{3/2}\,d_F$, where the dielectric term cancels two thirds of the Hall term, bounding the repulsive pressure by
\begin{equation}
\frac{F}{A}\le
\frac{\hbar c\,\alpha^{2}\,C_1 C_2}{8\pi^{2}\,d_{\mathrm{pk}}^{4}}
=\frac{8\pi^{2}}{81}\,\hbar c\,\alpha^{6}\,
\frac{(a^{*}_{\perp})^{4}\,n_1^{2}n_2^{2}}{C_1 C_2}.
\label{eq:Fpk-bound}
\end{equation}
Equation~\eqref{eq:Fpk-bound} is the magnitude counterpart of Eq.~\eqref{eq:dF-bound}: at fixed densities the maximal repulsive pressure decreases as $1/(C_1C_2)$. Moreover, dropping the negative dielectric term in Eq.~\eqref{eq:F-constrained} caps the repulsion at any separation by the Hall tail by a fraction $90\,\alpha^{2}C_1C_2/\pi^{4}$ of the perfect-metal pressure $\pi^{2}\hbar c/(240\,d^{4})$.

In the remainder of this work we discuss three examples: Landau levels originating from a parabolic band, a three-band Chern insulator with flat-bands, and twisted MoTe$_2$.

\textit{Example 1: Landau levels of parabolic bands.-} The bound~\eqref{eq:susc-bound} saturates for the integer quantum Hall state of a
two-dimensional free-electron gas in a transverse magnetic field
$B$. With the $\nu$ lowest Landau levels filled, the magnitude
of the Chern number equals $\nu$, and the areal density is
\begin{equation}
n_e=\frac{\nu}{hc}|eB|=\frac{m_e\,\omega_c\,|C|}{h},
\label{eq:LL-density}
\end{equation}
where $\omega_c=|eB|/(m_e c)$ is the cyclotron frequency. Inserting
\eqref{eq:LL-density} into the saturated bound and using
$e^2 m_e/\hbar^2=1/a_0$,
\begin{align}
\chitwoD(0)
&=\frac{2e^2|C|}{\hbar\omega_c},
\label{eq:LL-chi}
\end{align}
which agrees with the direct evaluation of the Landau-level
polarizability~\cite{Passos2026}. 
The crossover~\eqref{eq:dF-bound} for the Landau-level case follows by inserting
\eqref{eq:LL-chi}:
\begin{equation}
d_F=\dfrac{\sqrt{3}}{2}\,\frac{\chi(0)}{\alpha C}
=\sqrt{3}\,\frac{\hbar c}{\Eg},
\label{eq:LL-dcross}
\end{equation}
where we have used that the Landau gap is $\Eg=\hbar\omega_c$ and $e^2=\alpha\hbar c$. 
Because $\chi\propto|C|$ at fixed $\omega_c$, $d_F$ is independent of
$C$ at fixed field, while at fixed density $\chi\propto C^2$ and $d_F\propto|C|$
as in Eq.~\eqref{eq:dF-bound}. 
Numerically, $d_F=\sqrt{3}\,\hbar c/\Eg\simeq1.73\,\hbar c/\Eg$.
Beyond this distance the Casimir force
is repulsive.

We now recall that the long-distance expansion above is an approximation, and one can wonder how accurate this bound is for this example. In fact, the long-distance expansion overestimates the exact crossovers, and hence it is in this sense {overly conservative} compared to the exact value. This can be checked for the Landau level example by retaining the full frequency dependence of the Landau-level conductivities (Appendix~\ref{app:LLbenchmark}) in an expansion of the Casimir in orders of $\alpha \approx 10^{-2}$. This calculation leads to a Casimir integral that can be evaluated numerically, yielding 
${d_F^{\mathrm{ex}}}=1.383\,\hbar c/\Eg$. This value lies about $25\%$ below the
long-distance predictions, because at $\hbar\xi\sim\Eg$, when the long-distance approximation ceases to be controlled, the true $\sxx(\ii\xi)$ falls below
its linear-slope extrapolation $\xi\chitwoD(0)/4\pi$ and the attraction is weaker than
the truncation suggests. Hence, the bound overestimates attraction in the window of distances where it is valid.

To estimate this crossover distance we take a spinless Landau-level system at $B=10\ \mathrm{T}$ with the lowest level filled ($C=1$).  For a GaAs/AlGaAs two-dimensional electron gas we replace $(m_e,a_0)\to (m^{*},a^{*}_{\perp})$, and take $m^\ast=0.067m_e$, as appropriate for the GaAs conduction band, and a representative high-mobility density $n_e\simeq 2.4\times10^{11}\,{\rm cm}^{-2}$ \cite{Kriisa2019}.  
The cyclotron gap is $E_g=\hbar\omega_c=\frac{\hbar eB}{m^\ast}\simeq 17.3\,{\rm meV},$
and therefore $\frac{\hbar c}{E_g}\simeq 11.4\,\mu{\rm m}$.
In this case ${d_\times}=1.342\,\frac{\hbar c}{E_g}\simeq 15.3\,\mu{\rm m},$ 
and ${d_F}=\sqrt{3}\,\frac{\hbar c}{E_g}\simeq 19.8\,\mu{\rm m}$. These are in the $\mu$m range, the typical scale at which Casimir forces are observable~\cite{BKMM09}. We recall however that for insulating Chern plates the Casimir force is predicted to be much smaller than for metals, typically a factor $10^{-3}$~\cite{TseMacDonald2012}.

{\textit{Example 2: Ideal flat-bands.-}} %
To test how band flattening affects the onset of Casimir repulsion, we evaluated the
large-distance force sign-change distance for the three-band square-lattice Chern
model used by Passos and Souza~\cite{Passos2026}, originally introduced by
Yang et al.~\cite{YGS12}.  The occupied lowest band has
$C=3$ and its flatness can be tuned through a model parameter $\Delta$. At $\Delta=1$ the lowest band is maximally flat, while at $\Delta=3$ the bands are dispersive~\cite{Passos2026}. The Hall contribution to the large-distance interaction is fixed
topologically and  flattening can only enter through the longitudinal
polarizability $\chi^{2D}(0)$, which controls the attractive $d^{-5}$ term.  For
two identical sheets, the leading large-distance energy gives
\begin{equation}
  d_F(\Delta)
  =
  \frac{\sqrt{3}}{2}\,
  \frac{\chi^{2D}(0;\Delta)}{\alpha |C|},
  \label{eq:flatband-main-dF}
\end{equation}
where $d_F$ is the pressure sign-change distance.  The corresponding lower envelope
is obtained by replacing $\chi^{2D}(0;\Delta)$ by the Passos-Souza Chern
susceptibility bound for this model, calculated in Appendix \ref{app:flatband-casimir-benchmark}.  Figure~\ref{fig:flatband-main-distance} shows that the
actual onset distance follows $\chi^{2D}(0;\Delta)$: it is large for poorly
optimized bands, decreases as the Chern band is flattened and the susceptibility
approaches its bound, and then increases again for wide bands.  Flattening suppresses the attractive
longitudinal channel and moves the repulsive-force onset to shorter distances.

\textit{Example 3: Twisted MoTe$_2$ (t-MoTe$_2$).-}
To get a sense of the distances predicted by these bounds we
consider twisted MoTe$_2$, which realizes integer and
fractional Chern-insulating states at zero magnetic field~\cite{Cai2023,Zeng2023,Park2023}. The moir\'e density is much
lower than in an atomic-scale lattice Chern insulator: for twist angle
$\theta \simeq 3.7^{\circ}$, $a_m \simeq 5.5\,$nm and one carrier per
moir\'e cell gives $n_e \simeq 3.9 \times 10^{12}\,\mathrm{cm}^{-2}$.
The optical mass entering the bound is the bare hole mass of the
monolayer, $m^* = 0.62\, m_e$ [\onlinecite{Wu2019}]: the moir\'e
potential and interlayer tunneling are momentum independent and do not
renormalize the $f$-sum, so the flat moir\'e band carries a large
transport mass but an unrenormalized optical mass. Taking the integer
anomalous-Hall state with $C=1$, Eq.~\eqref{eq:dcross-identical}
becomes
\begin{equation}
d_{\times} \geq \frac{3}{\sqrt{20}\,\pi}
\frac{|C|}{\alpha\, a^{*}_{\perp} n_e}
\simeq 8.8\,\mu\mathrm{m},
\label{eq:dcross-tmote2}
\end{equation}
varying between $8$ and $18\,\mu$m over $n_e \simeq 3.9 \times 10^{12}\,\mathrm{cm}^{-2}$.
The thermodynamic charge gap of the $\nu=-1$ state measured by
nanoSQUID magnetometry is $\Eg \simeq 14\,$meV \cite{Redekop2024}, so that
$\hbar c/\Eg \simeq 14\,\mu$m and the crossover bound sits at
$d_{\times} \simeq 0.6\, \hbar c/\Eg$, in the same regime as the
Landau-level example.

An interesting, but so far purely academic, way to control the sign of the Casimir force: an attractive force corresponding to trivial insulator plates may be turned repulsive by entering a many-body phase with the correct Chern number signs~\cite{Cai2023,Zeng2023,Park2023}. This is in principle possible by tuning the gate voltage to trigger the many-body instability either to an anomalous Hall crystal, or to a fractional Chern insulator.

\textit{Discussion.-}
In this work we have bounded the extent to which Casimir repulsion can occur using quantum geometric bounds on DC responses. We have shown that this bound is a conservative bound since including the full finite-frequency response will weaken the attraction relative to its static extrapolation. Accordingly, the true crossover to repulsion occurs at shorter, more experimentally accessible separations than the bound indicates, as discussed in Fig.~\ref{fig:LLcrossover}. Using a three-band Chern insulator example we have shown that the crossover distance is shortest for plates that saturate the susceptibility bound, as shown by Fig.~\ref{fig:flatband-main-distance}. We end by discussing several consequences of our results in the context of previous works, and an outlook.

First, we recall that increasing the Chern number was suggested as a way to enhance repulsion when $C_1C_2>0$, because the repulsive term scales with the product of Chern numbers, $C^2$~\cite{TseMacDonald2012,Rodriguez-Lopez2014}. Our results put a geometric bound on this benefit: the attractive term bound scales as $C^4$, which dominates the repulsive $C^2$ at large Chern numbers. 
Consistently, Eq.~\eqref{eq:main-bound} pushes the crossover to larger, experimentally inaccessible separations. For $C_1C_2<0$ both terms in Eq.~\eqref{eq:E-combined} are attractive, and the geometric bound then implies enhanced attraction at all separations.

Second, because flat-bands with non-zero Chern number usually exhibit a parameter regime that saturates geometric bounds, our results rationalize why Landau levels and flat Chern bands are judicious choices to maximize the range where repulsion can be found by minimizing the crossover distance. Within realistic two-dimensional materials, such as twisted MoTe$_2$, this distance falls in the observable $\mu$m range. Hence, our results highlight the geometric origin of previously known results on Casimir repulsion in Landau levels~\cite{TseMacDonald2012}. 

Third, the bounds quantify the experimental challenge to measure repulsion in insulators. The repulsive pressure never exceeds a universal fraction $90\,\alpha^{2}C^{2}/\pi^{4}\simeq5\times10^{-5}\,C^{2}$ of the perfect-metal attraction at the same separation, see Eq.~\eqref{eq:F-constrained}.
Hence, our bounds show that the magnitude of the repulsive force is necessarily smaller than attraction seen in the metallic case, consistent with earlier estimates~\cite{TseMacDonald2012,Rodriguez-Lopez2014}.
With the present bounds we have geometrically reinterpreted this result: since $\chi(0)$ diverges for metals, the Casimir force for metallic plates, with or without Chern bands, is large in magnitude and attractive in sign.

While we have focused on two-dimensional systems, it is likely possible to derive bounds on three-dimensional topological metals based on Ref.~\cite{Passos2026}. 
For example it would be interesting to geometrically bound 
the Casimir repulsion predicted for topological semimetals~\cite{Grushin2012,Wilson2015}. 

In short, our work adds new quantum-geometric bounds that restrict the observation of Casimir repulsion in real materials. Together with previously known theorems~\cite{Lam97,Kenneth2006,B06,RK09,RE09,Rosa2010}, these bounds must be taken into account when designing novel ways to achieve the elusive Casimir repulsion in vacuum.\\

\textit{Acknowledgments.-}
I thank Daniel J. Passos for conversations that inspired this work, and him and Ivo Souza for insightful discussions.
In preparing this manuscript I used OpenAI's 5.5 model and Anthropic's Claude Fable, Opus 4.7 and 4.8 to develop mathematical derivations, figures and text editing. 
The models were capable of deriving the bounds after prompting them to correct some factors. 
The author re-derived the results, and remains responsible for the content. The Mathematica notebooks, together with the code that reproduces Fig. 2 and 3 are available in the Zenodo repository \cite{grushin_zenodo_2026}. The author is supported by the European Research Council (ERC) Consolidator grant under grant agreement No. 101042707 (TOPOMORPH).\\

\bibliography{bounds}

@Article{Passos2026,
	title={{Electronic bounds in magnetic crystals}},
	author={Daniel Passos and Ivo Souza},
	journal={SciPost Phys.},
	volume={20},
	pages={093},
	year={2026},
	publisher={SciPost},
	doi={10.21468/SciPostPhys.20.3.093},
	url={https://scipost.org/10.21468/SciPostPhys.20.3.093},
}

@article{provost80,
  title = {Riemannian structure on manifolds of quantum states},
  author = {Provost, J. P. and Vallee, G.},
  journal = {Commun. Math. Phys.},
  volume = {76},
  pages = {289},
  year = {1980},
  doi = {10.1007/BF02193559}
}

@article{berry84,
  author       = "M. V. Berry",
  title        = "Quantal phase factors accompanying adiabatic changes",
  journal      = "Proc. R. Soc. Lond. A",
  volume       = 392,
  pages        = 45,
  year         = 1984,
  doi          ={10.1098/rspa.1984.0023}
  }

@article{yu-npjquantummater25,
	title = {Quantum geometry in quantum materials},
	author = {Jiabin Yu and B. Andrei Bernevig and Raquel Queiroz
                  and Enrico Rossi and Päivi Törmä and Bohm-Jung Yang},
	journal = {npj Quantum Mater.},
	volume = {10},
	pages = {101},
	year = {2025},
	doi = {10.1038/s41535-025-00801-3}
}

@misc{verma2025quantumgeometryrevisitingelectronic,
      title={Quantum Geometry: Revisiting electronic scales in quantum matter}, 
      author={Nishchhal Verma and Philip J. W. Moll and Tobias Holder and Raquel Queiroz},
      year={2025},
      eprint={2504.07173},
      archivePrefix={arXiv},
      url={https://arxiv.org/abs/2504.07173},
      doi = {10.48550/arXiv.2504.07173}
}

@misc{Gao2025quantumgeometry,
      title={Quantum Geometry Phenomena in Condensed Matter Systems},
      author={Anyuan Gao and Naoto Nagaosa and Ni N and Su-Yang Xu},
      year={2025},
      eprint={2508.00469},
      archivePrefix={arXiv},
      url={https://arxiv.org/abs/2508.00469},
      doi={10.48550/arXiv.2508.00469}
}

@article{roy-prb14,
  title = {Band geometry of fractional topological insulators},
  author = {Roy, Rahul},
  journal = {Phys. Rev. B},
  volume = {90},
  issue = {16},
  pages = {165139},
  numpages = {7},
  year = {2014},
  month = {Oct},
  publisher = {American Physical Society},
  doi = {10.1103/PhysRevB.90.165139},
  url = {https://link.aps.org/doi/10.1103/PhysRevB.90.165139}
}

@article{peotta-natcomms15,
	title = {Superfluidity in topologically nontrivial flat bands},
	author = {S. Peotta and P. T\"orm\"a},
	journal = {Nat. Communs.},
	volume = {6},
	pages = {8944},
	year = {2015},
	doi = {10.1038/ncomms9944},
	url = {}
}

@article{ozawa-prb21,
  title = {{Relations between topology and the quantum metric for
                  Chern insulators}},
  author = {Ozawa, Tomoki and Mera, Bruno},
  journal = {Phys. Rev. B},
  volume = {104},
  issue = {4},
  pages = {045103},
  numpages = {13},
  year = {2021},
  month = {Jul},
  publisher = {American Physical Society},
  doi = {10.1103/PhysRevB.104.045103},
  url = {https://link.aps.org/doi/10.1103/PhysRevB.104.045103}
}

@article{souza-prb00,
  title = {Polarization and localization in insulators: Generating function approach},
  author = {Souza, Ivo and Wilkens, Tim and Martin, Richard M.},
  journal = {Phys. Rev. B},
  volume = {62},
  issue = {3},
  pages = {1666},
  numpages = {0},
  year = {2000},
  month = {Jul},
  publisher = {American Physical Society},
  doi = {10.1103/PhysRevB.62.1666},
  url = {https://link.aps.org/doi/10.1103/PhysRevB.62.1666}
}

@article{aebischer-prl01,
  title = {{Dielectric Catastrophe at the Mott Transition}},
  author = {Aebischer, C. and Baeriswyl, D. and Noack, R. M.},
  journal = {Phys. Rev. Lett.},
  volume = {86},
  issue = {3},
  pages = {468},
  numpages = {0},
  year = {2001},
  month = {Jan},
  publisher = {American Physical Society},
  doi = {10.1103/PhysRevLett.86.468},
  url = {https://link.aps.org/doi/10.1103/PhysRevLett.86.468}
}

@book{martin-book04,
  author       = {R. M. Martin},
  title        = {{Electronic Structure: Basic Theory and Practical Methods}},
  publisher    = {Cambridge},
  edition      = {1st edition},
  isbn         = {0521782856},
  year         = 2004}

@article{onishi-prx24,
  title = {Fundamental Bound on Topological Gap},
  author = {Onishi, Yugo and Fu, Liang},
  journal = {Phys. Rev. X},
  volume = {14},
  issue = {1},
  pages = {011052},
  numpages = {12},
  year = {2024},
  month = {Mar},
  publisher = {American Physical Society},
  doi = {10.1103/PhysRevX.14.011052},
  url = {https://link.aps.org/doi/10.1103/PhysRevX.14.011052}
}

@article{komissarov-natcomms24,
   title = {The quantum geometric origin of capacitance in insulators}, 
   author = {Ilia Komissarov and Tobias Holder and Raquel Queiroz},
   journal = {Nature Commun.},
   volume = {15},
   pages = {4621},
   year = {2024},
   doi = {10.1038/s41467-024-48808-x}
}

@article{onishi-prb24,
  title = {Universal relation between energy gap and dielectric constant},
  author = {Onishi, Yugo and Fu, Liang},
  journal = {Phys. Rev. B},
  volume = {110},
  issue = {15},
  pages = {155107},
  numpages = {7},
  year = {2024},
  month = {Oct},
  publisher = {American Physical Society},
  doi = {10.1103/PhysRevB.110.155107},
  url = {https://link.aps.org/doi/10.1103/PhysRevB.110.155107}
}

@article{verma-pnas25,
   title={Instantaneous response and quantum geometry of insulators},
   volume={122},
   pages={e2405837122},
   DOI={10.1073/pnas.2405837122},
   journal={Proc. Natl. Acad. Sci. U.S.A.},
   author={Verma, Nishchhal and Queiroz, Raquel},
   year={2025},
   month=dec }

@article{souza-scipost25,
	title={{Optical bounds on many-electron localization}},
	author={Ivo Souza and Richard M.  Martin and Massimiliano Stengel},
	journal={SciPost Phys.},
	volume={18},
	pages={127},
	year={2025},
	publisher={SciPost},
	doi={10.21468/SciPostPhys.18.4.127},
	url={https://scipost.org/10.21468/SciPostPhys.18.4.127},
}

@article{onishi-prr25,
  title = {Quantum weight: A fundamental property of quantum many-body systems},
  author = {Onishi, Yugo and Fu, Liang},
  journal = {Phys. Rev. Res.},
  volume = {7},
  issue = {2},
  pages = {023158},
  numpages = {11},
  year = {2025},
  month = {May},
  publisher = {American Physical Society},
  doi = {10.1103/PhysRevResearch.7.023158},
  url = {https://link.aps.org/doi/10.1103/PhysRevResearch.7.023158}
}

@article{Woods2016,
  title = {Materials perspective on Casimir and van der Waals interactions},
  author = {Woods, L. M. and Dalvit, D. A. R. and Tkatchenko, A. and Rodriguez-Lopez, P. and Rodriguez, A. W. and Podgornik, R.},
  journal = {Rev. Mod. Phys.},
  volume = {88},
  issue = {4},
  pages = {045003},
  numpages = {48},
  year = {2016},
  month = {Nov},
  publisher = {American Physical Society},
  doi = {10.1103/RevModPhys.88.045003},
  url = {https://link.aps.org/doi/10.1103/RevModPhys.88.045003}
}

@article{Zhao2009,
  title = {Repulsive Casimir Force in Chiral Metamaterials},
  author = {Zhao, R. and Zhou, J. and Koschny, Th. and Economou, E. N. and Soukoulis, C. M.},
  journal = {Phys. Rev. Lett.},
  volume = {103},
  issue = {10},
  pages = {103602},
  numpages = {4},
  year = {2009},
  month = {Sep},
  publisher = {American Physical Society},
  doi = {10.1103/PhysRevLett.103.103602},
  url = {https://link.aps.org/doi/10.1103/PhysRevLett.103.103602}
}

@article{Kenneth2006,
  title = {Opposites Attract: A Theorem about the Casimir Force},
  author = {Kenneth, Oded and Klich, Israel},
  journal = {Phys. Rev. Lett.},
  volume = {97},
  issue = {16},
  pages = {160401},
  numpages = {4},
  year = {2006},
  month = {Oct},
  publisher = {American Physical Society},
  doi = {10.1103/PhysRevLett.97.160401},
  url = {https://link.aps.org/doi/10.1103/PhysRevLett.97.160401}
}

@article{Rosa2010,
  title = {Repulsive Casimir force: Sufficient conditions},
  author = {Rosa, Luigi and Lambrecht, Astrid},
  journal = {Phys. Rev. D},
  volume = {82},
  issue = {6},
  pages = {065025},
  numpages = {6},
  year = {2010},
  month = {Sep},
  publisher = {American Physical Society},
  doi = {10.1103/PhysRevD.82.065025},
  url = {https://link.aps.org/doi/10.1103/PhysRevD.82.065025}
}

@article{Cas48,
  author  = {Casimir, H. B. G.},
  title   = {On the attraction between two perfectly conducting plates},
  journal = {Proc. Kon. Neder. Akad. Wet.},
  volume  = {51},
  pages   = {793--795},
  year    = {1948},
  url     = {https://www.dwc.knaw.nl/DL/publications/PU00018547.pdf}
}

@book{BKMM09,
  author    = {Bordag, Michael and Klimchitskaya, Galina L. and Mohideen, Umar and Mostepanenko, Vladimir M.},
  title     = {Advances in the Casimir Effect},
  publisher = {Oxford University Press},
  year      = {2009},
  doi       = {10.1093/acprof:oso/9780199238743.001.0001},
  url       = {https://doi.org/10.1093/acprof:oso/9780199238743.001.0001}
}

@article{RE09,
  author  = {Rahi, Sahand Jamal and Emig, Thorsten and Graham, Noah and Jaffe, Robert L. and Kardar, Mehran},
  title   = {Scattering theory approach to electrodynamic Casimir forces},
  journal = {Phys. Rev. D},
  volume  = {80},
  pages   = {085021},
  year    = {2009},
  doi     = {10.1103/PhysRevD.80.085021},
  url     = {https://doi.org/10.1103/PhysRevD.80.085021}
}

@article{DLP61,
  author  = {Dzyaloshinskii, I. E. and Lifshitz, E. M. and Pitaevskii, L. P.},
  title   = {The general theory of van der Waals forces},
  journal = {Advances in Physics},
  volume  = {10},
  pages   = {165--209},
  year    = {1961},
  doi     = {10.1080/00018736100101281},
  url     = {https://doi.org/10.1080/00018736100101281}
}

@article{RK09,
  author  = {Rahi, Sahand Jamal and Kardar, Mehran and Emig, Thorsten},
  title   = {Constraints on Stable Equilibria with Fluctuation-Induced Forces},
  journal = {Phys. Rev. Lett.},
  volume  = {105},
  pages   = {070404},
  year    = {2010},
  doi     = {10.1103/PhysRevLett.105.070404},
  url     = {https://doi.org/10.1103/PhysRevLett.105.070404}
}

@article{B06,
  author  = {Bachas, C. P.},
  title   = {Comment on the sign of the Casimir force},
  journal = {Journal of Physics A: Mathematical and Theoretical},
  volume  = {40},
  pages   = {9089--9096},
  year    = {2007},
  doi     = {10.1088/1751-8113/40/30/S23},
  url     = {https://doi.org/10.1088/1751-8113/40/30/S23}
}

@article{Lam97,
  author = {A. Lambrecht and M.-T. Jaekel and S. Reynaud},
  title = {The Casimir force for passive mirrors},
  journal = {Physics Letters A},
  year = {1997},
  volume = {225},
  pages = {188--194}
}

@article{Wu2019,
  author = {Wu, Fengcheng and Lovorn, Timothy and Tutuc, Emanuel and Martin, Ivar and MacDonald, A. H.},
  title = {Topological Insulators in Twisted Transition Metal Dichalcogenide Homobilayers},
  journal = {Physical Review Letters},
  volume = {122},
  pages = {086402},
  year = {2019},
  doi = {10.1103/PhysRevLett.122.086402}
}

@article{Redekop2024,
  author = {Redekop, Evgeny and Zhang, Canxun and Park, Heonjoon and Cai, Jiaqi and Anderson, Eric and Sheekey, Owen and Arp, Trevor and Babikyan, Grigory and Salters, Samuel and Watanabe, Kenji and Taniguchi, Takashi and Huber, M. E. and Xu, Xiaodong and Young, A. F.},
  title = {Direct Magnetic Imaging of Fractional Chern Insulators in Twisted {MoTe$_2$} with a Superconducting Sensor},
  journal = {Nature},
  volume = {635},
  pages = {584--589},
  year = {2024},
  doi = {10.1038/s41586-024-08153-x},
  eprint = {2405.10269},
  archivePrefix = {arXiv},
  primaryClass = {cond-mat.mes-hall}
}

@article{B74,
  author  = {Boyer, Timothy H.},
  title   = {Van der Waals forces and zero-point energy for dielectric and permeable materials},
  journal = {Phys. Rev. A},
  volume  = {9},
  pages   = {2078--2084},
  year    = {1974},
  doi     = {10.1103/PhysRevA.9.2078},
  url     = {https://doi.org/10.1103/PhysRevA.9.2078}
}

@article{Rosa08,
  author  = {Rosa, Felipe S. S. and Dalvit, Diego A. R. and Milonni, Peter W.},
  title   = {Casimir-Lifshitz Theory and Metamaterials},
  journal = {Phys. Rev. Lett.},
  volume  = {100},
  pages   = {183602},
  year    = {2008},
  doi     = {10.1103/PhysRevLett.100.183602},
  url     = {https://doi.org/10.1103/PhysRevLett.100.183602}
}

@article{LMR10,
  author  = {Levin, Michael and McCauley, Alexander P. and Rodriguez, Alejandro W. and Reid, M. T. Homer and Johnson, Steven G.},
  title   = {Casimir Repulsion between Metallic Objects in Vacuum},
  journal = {Phys. Rev. Lett.},
  volume  = {105},
  pages   = {090403},
  year    = {2010},
  doi     = {10.1103/PhysRevLett.105.090403},
  url     = {https://doi.org/10.1103/PhysRevLett.105.090403}
}

@article{BV00,
  author  = {Bordag, M. and Vassilevich, D. V.},
  title   = {Casimir force between Chern-Simons surfaces},
  journal = {Physics Letters A},
  volume  = {268},
  pages   = {75--80},
  year    = {2000},
  doi     = {10.1016/S0375-9601(00)00167-7},
  url     = {
https://doi.org/10.1016/S0375-9601%2800%2900159-6}
}

@article{GC11,
  author  = {Grushin, Adolfo G. and Cortijo, Alberto},
  title   = {Tunable Casimir Repulsion with Three-Dimensional Topological Insulators},
  journal = {Phys. Rev. Lett.},
  volume  = {106},
  pages   = {020403},
  year    = {2011},
  doi     = {10.1103/PhysRevLett.106.020403},
  url     = {https://doi.org/10.1103/PhysRevLett.106.020403}
}

@article{GRC11,
  author  = {Grushin, Adolfo G. and Rodriguez-Lopez, Pablo and Cortijo, Alberto},
  title   = {Effect of finite temperature and uniaxial anisotropy on the Casimir effect with three-dimensional topological insulators},
  journal = {Phys. Rev. B},
  volume  = {84},
  pages   = {045119},
  year    = {2011},
  doi     = {10.1103/PhysRevB.84.045119},
  url     = {https://doi.org/10.1103/PhysRevB.84.045119}
}

@article{YGS12,
  author  = {Yang, Shengyuan A. and Gu, Zheng-Cheng and Sun, Kai and Das Sarma, S.},
  title   = {Topological flat band models with arbitrary Chern numbers},
  journal = {Phys. Rev. B},
  volume  = {86},
  pages   = {241112},
  year    = {2012},
  doi     = {10.1103/PhysRevB.86.241112},
  url     = {https://doi.org/10.1103/PhysRevB.86.241112}
}

@article{IKJ13,
  author  = {Intravaia, Francesco and Koev, Stephan and Jung, I. W. and Talin, A. Alec and Davids, Paul S. and Decca, Ricardo S. and Aksyuk, Vladimir A. and Dalvit, Diego A. R. and Lopez, Daniel},
  title   = {Strong Casimir force reduction through metallic surface nanostructuring},
  journal = {Nature Communications},
  volume  = {4},
  pages   = {2515},
  year    = {2013},
  doi     = {10.1038/ncomms3515},
  url     = {https://doi.org/10.1038/ncomms3515}
}

@book{LLv8,
  author    = {Landau, L. D. and Lifshitz, E. M.},
  title     = {Electrodynamics of Continuous Media},
  publisher = {Pergamon Press},
  address   = {Oxford},
  edition   = {2},
  year      = {1984},
  isbn      = {9780080302751},
  url       = {https://www.sciencedirect.com/book/9780080302751/electrodynamics-of-continuous-media}
}

@book{DG02,
  author    = {Dressel, Martin and Gruner, George},
  title     = {Electrodynamics of Solids: Optical Properties of Electrons in Matter},
  publisher = {Cambridge University Press},
  year      = {2002},
  doi       = {10.1017/CBO9780511606168},
  url       = {https://doi.org/10.1017/CBO9780511606168}
}

@article{Lifshitz1956,
  author  = {Lifshitz, E. M.},
  title   = {The Theory of Molecular Attractive Forces between Solids},
  journal = {Soviet Physics JETP},
  volume  = {2},
  pages   = {73},
  year    = {1956}
}

@article{TseMacDonald2012,
  author  = {Tse, Wang-Kong and MacDonald, A. H.},
  title   = {Quantized Casimir Force},
  journal = {Physical Review Letters},
  volume  = {109},
  pages   = {236806},
  year    = {2012},
  doi     = {10.1103/PhysRevLett.109.236806},
  url     = {https://doi.org/10.1103/PhysRevLett.109.236806}
}

@article{FKV2018,
  author        = {Fialkovsky, I. and Khusnutdinov, N. and Vassilevich, D.},
  title         = {Quest for Casimir Repulsion between Chern-Simons Surfaces},
  journal       = {Physical Review B},
  volume        = {97},
  pages         = {165432},
  year          = {2018},
  doi           = {10.1103/PhysRevB.97.165432},
  url           = {https://doi.org/10.1103/PhysRevB.97.165432},
  eprint        = {1802.06598},
  archivePrefix = {arXiv},
  primaryClass  = {cond-mat.mes-hall}
}

@article{Cai2023,
  author  = {Cai, Jiaqi and Anderson, Eric and Wang, Chang and Zhang, Xiaodong and Liu, Xiaoyu and Holtzmann, William and Zhang, Yiran and Fan, Feng-Ren and Taniguchi, Takashi and Watanabe, Kenji and Ran, Ying and Cao, Ting and Xiao, Di and Fu, Liang and Yao, Wang and Cobden, David and Xu, Xiaodong},
  title   = {Signatures of fractional quantum anomalous Hall states in twisted MoTe2},
  journal = {Nature},
  volume  = {622},
  pages   = {63--68},
  year    = {2023},
  doi     = {10.1038/s41586-023-06289-w}
}

@article{Zeng2023,
  author  = {Zeng, Yihang and Xia, Zhengchao and Kang, Kaifei and Zhu, Jiacheng and Kn{\"u}ppel, Patrick and Vaswani, Chirag and Watanabe, Kenji and Taniguchi, Takashi and Mak, Kin Fai and Shan, Jie},
  title   = {Thermodynamic evidence of fractional Chern insulator in moire MoTe2},
  journal = {Nature},
  volume  = {622},
  pages   = {69--73},
  year    = {2023},
  doi     = {10.1038/s41586-023-06452-3}
}

@article{grushin_zenodo_2026,
	title        = {{Quantum-geometric bounds on Casimir repulsion}},
	author       = {Grushin, Adolfo G.},
	year         = 2026,
	journal      = {Zenodo},
	doi          = {10.5281/zenodo.21997247}
}

@article{Park2023,
  author  = {Park, Heonjoon and Cai, Jiaqi and Anderson, Eric and Wang, Chang and Zhang, Xiaodong and Liu, Xiaoyu and Holtzmann, William and Zhang, Yiran and Fan, Feng-Ren and Taniguchi, Takashi and Watanabe, Kenji and Ran, Ying and Cao, Ting and Xiao, Di and Fu, Liang and Yao, Wang and Cobden, David and Xu, Xiaodong},
  title   = {Observation of fractionally quantized anomalous Hall effect},
  journal = {Nature},
  volume  = {622},
  pages   = {74--79},
  year    = {2023},
  doi     = {10.1038/s41586-023-06536-0}
}

@article{RLKortKamp2017,
  author  = {Rodriguez-Lopez, P. and Kort-Kamp, W. J. M. and Dalvit, D. A. R. and Woods, L. M.},
  title   = {Casimir Force Phase Transitions in the Graphene Family},
  journal = {Nature Communications},
  volume  = {8},
  pages   = {14699},
  year    = {2017},
  doi     = {10.1038/ncomms14699},
  url     = {https://doi.org/10.1038/ncomms14699}
}

@article{Lam08,
  author = {A. Lambrecht and I. Pirozhenko},
  title = {Casimir force between dissimilar mirrors and the role of the surface plasmons},
  journal = {Physical Review A},
  year = {2008},
  volume = {78},
  pages = {062102},
  doi = {10.1103/PhysRevA.78.062102},
  url = {https://doi.org/10.1103/PhysRevA.78.062102}
}

@article{Pir09,
  author = {I. Pirozhenko and A. Lambrecht},
  title = {Repulsive Casimir forces and the role of surface modes},
  journal = {Physical Review A},
  year = {2009},
  volume = {80},
  pages = {042510},
  doi = {10.1103/PhysRevA.80.042510},
  url = {https://doi.org/10.1103/PhysRevA.80.042510}
}

@article{Zwo10,
  author = {P. J. van Zwol and G. Palasantzas},
  title = {Repulsive Casimir forces between solid materials with high-refractive-index intervening liquids},
  journal = {Physical Review A},
  year = {2010},
  volume = {81},
  pages = {062502},
  doi = {10.1103/PhysRevA.81.062502},
  url = {https://doi.org/10.1103/PhysRevA.81.062502}
}

@article{Mun09b,
  author = {J. N. Munday and F. Capasso and V. A. Parsegian},
  title = {Measured long-range repulsive Casimir--Lifshitz forces},
  journal = {Nature},
  year = {2009},
  volume = {457},
  pages = {170--173},
  doi = {10.1038/nature07610},
  url = {https://doi.org/10.1038/nature07610}
}

@article{Rodriguez-Lopez2014,
  title = {Repulsive Casimir Effect with Chern Insulators},
  author = {Rodriguez-Lopez, Pablo and Grushin, Adolfo G.},
  journal = {Phys. Rev. Lett.},
  volume = {112},
  issue = {5},
  pages = {056804},
  numpages = {5},
  year = {2014},
  month = {Feb},
  publisher = {American Physical Society},
  doi = {10.1103/PhysRevLett.112.056804},
  url = {https://link.aps.org/doi/10.1103/PhysRevLett.112.056804}
}

@article{Metalidis2002,
  title = {Magnetic Casimir effect},
  author = {Metalidis, G. and Bruno, P.},
  journal = {Phys. Rev. A},
  volume = {66},
  issue = {6},
  pages = {062102},
  numpages = {10},
  year = {2002},
  month = {Dec},
  publisher = {American Physical Society},
  doi = {10.1103/PhysRevA.66.062102},
  url = {https://link.aps.org/doi/10.1103/PhysRevA.66.062102}
}

@article{Metalidis2010,
  title = {Magnetic anisotropy due to the Casimir effect},
  author = {Metalidis, G. and Bruno, P.},
  journal = {Phys. Rev. A},
  volume = {81},
  issue = {2},
  pages = {022123},
  numpages = {6},
  year = {2010},
  month = {Feb},
  publisher = {American Physical Society},
  doi = {10.1103/PhysRevA.81.022123},
  url = {https://link.aps.org/doi/10.1103/PhysRevA.81.022123}
}

@article{Wilson2015,
  title = {Repulsive Casimir force between Weyl semimetals},
  author = {Wilson, Justin H. and Allocca, Andrew A. and Galitski, Victor},
  journal = {Phys. Rev. B},
  volume = {91},
  issue = {23},
  pages = {235115},
  numpages = {5},
  year = {2015},
  month = {Jun},
  publisher = {American Physical Society},
  doi = {10.1103/PhysRevB.91.235115},
  url = {https://link.aps.org/doi/10.1103/PhysRevB.91.235115}
}

@article{Grushin2012,
  title = {Consequences of a condensed matter realization of Lorentz-violating QED in Weyl semi-metals},
  author = {Grushin, Adolfo G.},
  journal = {Phys. Rev. D},
  volume = {86},
  issue = {4},
  pages = {045001},
  numpages = {7},
  year = {2012},
  month = {Aug},
  publisher = {American Physical Society},
  doi = {10.1103/PhysRevD.86.045001},
  url = {https://link.aps.org/doi/10.1103/PhysRevD.86.045001}
}

@article{Farias2020,
  title = {Casimir force between Weyl semimetals in a chiral medium},
  author = {Farias, M. Bel\'en and Zyuzin, Alexander A. and Schmidt, Thomas L.},
  journal = {Phys. Rev. B},
  volume = {101},
  issue = {23},
  pages = {235446},
  numpages = {9},
  year = {2020},
  month = {Jun},
  publisher = {American Physical Society},
  doi = {10.1103/PhysRevB.101.235446},
  url = {https://link.aps.org/doi/10.1103/PhysRevB.101.235446}
}

@article{Kriisa2019,
	author = {Kriisa, A. and Samaraweera, R. L. and Heimbeck, M. S. and Everitt, H. O. and Reichl, C. and Wegscheider, W. and Mani, R. G.},
	date = {2019/02/20},
	doi = {10.1038/s41598-019-39186-2},
	id = {Kriisa2019},
	isbn = {2045-2322},
	journal = {Scientific Reports},
	number = {1},
	pages = {2409},
	title = {Cyclotron resonance in the high mobility GaAs/AlGaAs 2D electron system over the microwave, mm-wave, and terahertz- bands},
	url = {https://doi.org/10.1038/s41598-019-39186-2},
	volume = {9},
	year = {2019}}

@article{Vassilevich2020,
author = {Vassilevich, Dmitri},
title = {On the (im)possibility of Casimir repulsion between Chern-Simons surfaces},
journal = {Modern Physics Letters A},
volume = {35},
number = {03},
pages = {2040017},
year = {2020},
doi = {10.1142/S0217732320400179},
URL = { 
        https://doi.org/10.1142/S0217732320400179
}
}

@article{Rodriguez-Lopez2011,
  title = {Casimir repulsion between topological insulators in the diluted regime},
  author = {Rodriguez-Lopez, Pablo},
  journal = {Phys. Rev. B},
  volume = {84},
  issue = {16},
  pages = {165409},
  numpages = {7},
  year = {2011},
  month = {Oct},
  publisher = {American Physical Society},
  doi = {10.1103/PhysRevB.84.165409},
  url = {https://link.aps.org/doi/10.1103/PhysRevB.84.165409}
}

@article{Milton_2012,
doi = {10.1088/1751-8113/45/37/374006},
url = {https://doi.org/10.1088/1751-8113/45/37/374006},
year = {2012},
month = {sep},
publisher = {IOP Publishing},
volume = {45},
number = {37},
pages = {374006},
author = {Milton, Kimball A and Abalo, E K and Parashar, Prachi and Pourtolami, Nima and Brevik, Iver and Ellingsen, Simen A},
title = {Repulsive Casimir and Casimir–Polder forces},
journal = {Journal of Physics A: Mathematical and Theoretical}
}

@article{Morimoto2009,
  title = {Optical Hall Conductivity in Ordinary and Graphene Quantum Hall Systems},
  author = {Morimoto, Takahiro and Hatsugai, Yasuhiro and Aoki, Hideo},
  journal = {Phys. Rev. Lett.},
  volume = {103},
  issue = {11},
  pages = {116803},
  numpages = {4},
  year = {2009},
  month = {Sep},
  publisher = {American Physical Society},
  doi = {10.1103/PhysRevLett.103.116803},
  url = {https://link.aps.org/doi/10.1103/PhysRevLett.103.116803}
}

@article{Zhao2010,
  title = {Comparison of chiral metamaterial designs for repulsive Casimir force},
  author = {Zhao, R. and Koschny, Th. and Economou, E. N. and Soukoulis, C. M.},
  journal = {Phys. Rev. B},
  volume = {81},
  issue = {23},
  pages = {235126},
  numpages = {5},
  year = {2010},
  month = {Jun},
  publisher = {American Physical Society},
  doi = {10.1103/PhysRevB.81.235126},
  url = {https://link.aps.org/doi/10.1103/PhysRevB.81.235126}
}

@article{Deng2011,
  title = {Attractive-repulsive transition of the Casimir force between anisotropic plates},
  author = {Deng, Gang and Liu, Zhong-Zhu and Luo, Jun},
  journal = {Phys. Rev. A},
  volume = {78},
  issue = {6},
  pages = {062111},
  numpages = {5},
  year = {2008},
  month = {Dec},
  publisher = {American Physical Society},
  doi = {10.1103/PhysRevA.78.062111},
  url = {https://link.aps.org/doi/10.1103/PhysRevA.78.062111}
}

@article{Chen2011,
  title = {Casimir interaction between topological insulators with finite surface band gap},
  author = {Chen, Liang and Wan, Shaolong},
  journal = {Phys. Rev. B},
  volume = {84},
  issue = {7},
  pages = {075149},
  numpages = {6},
  year = {2011},
  month = {Aug},
  publisher = {American Physical Society},
  doi = {10.1103/PhysRevB.84.075149},
  url = {https://link.aps.org/doi/10.1103/PhysRevB.84.075149}
}

\setcounter{secnumdepth}{5}
\renewcommand{\theparagraph}{\bf \thesubsubsection.\arabic{paragraph}}

\renewcommand{\thefigure}{S\arabic{figure}}
\setcounter{figure}{0}

\appendix

\begin{widetext}
\newpage
\begin{center}
{\bf Supplemental Materials for ``Quantum-geometric bounds on Casimir repulsion''}
\end{center}

\appendix

\section{Explicit derivation of the leading large-distance energy}
\label{app:derivation}

This appendix fills in the steps between Eqs.~\eqref{eq:lifshitz} and \eqref{eq:E-combined} of the main text. We
(i) linearize the Tse--MacDonald reflection coefficients in the sheet conductivity,
(ii) expand the Lifshitz logarithm to leading order, (iii) evaluate the trace
$\mathrm{Tr}(R_1R_2)$ for a gapped Chern insulator, and (iv) perform the phase-space
integrals to obtain the two terms of Eq.~\eqref{eq:E-combined} .

\subsection{Linearization of the Tse--MacDonald reflection coefficients}
\label{app:linearization}

The reflection matrix of a single 2D sheet in vacuum, computed to all orders in the
local conductivity, is given by Tse and MacDonald~\cite{TseMacDonald2012}. In the
linear-polarization ($xy$) basis, with the Gaussian-normalized dimensionless
conductivity $\tilde\sigma_{ij}\equiv\sigma_{ij}/c$ and
$\lambda=\cos\theta=k_z/(\omega/c)$,
\begin{align}
r_{xx}&=-\frac{2\pi}{\mathcal R}
\left[\frac{\tilde\sigma_{xx}}{\lambda}
+2\pi\big(\tilde\sigma_{xx}^2+\tilde\sigma_{xy}^2\big)\right],
\label{eq:app-TMxx}\\
r_{yy}&=-\frac{2\pi}{\mathcal R}
\left[\tilde\sigma_{xx}\,\lambda
+2\pi\big(\tilde\sigma_{xx}^2+\tilde\sigma_{xy}^2\big)\right],
\label{eq:app-TMyy}\\
r_{xy}&=-r_{yx}=-\frac{2\pi}{\mathcal R}\,\tilde\sigma_{xy},
\label{eq:app-TMxy}
\end{align}
with the common denominator
\begin{equation}
\mathcal R=1+2\pi\tilde\sigma_{xx}\!\left(\lambda+\frac{1}{\lambda}\right)
+4\pi^2\big(\tilde\sigma_{xx}^2+\tilde\sigma_{xy}^2\big).
\label{eq:app-TMdenom}
\end{equation}
These are exact within the local (frequency-only) conductivity model.  The normal 
wavevector in this real-frequency formula continues as
$k_z\to i\kappa$ when $\omega\to i\xi$, with
$\kappa=\sqrt{q^2+\xi^2/c^2}$.  Thus the variable introduced later in
Eq.~\eqref{eq:app-xdef}, $x=\xi/(c\kappa)$, is related to the
Tse--MacDonald variable by
$\lambda=k_z/(\omega/c)\to i\kappa/(i\xi/c)=1/x$.  We keep $\lambda$
in this subsection to make the comparison with the notation of
Tse and MacDonald~\cite{TseMacDonald2012} transparent.

At large distance the dominant frequencies are small (Sec.~\ref{app:logexpand}), so
we can expand each sheet conductivity as a function of frequency as
\begin{align}
\sxy(\ii\xi)&=C_a\,\frac{e^2}{h}+\mathcal{O}\!\left(\frac{\hbar^2\xi^2}{\Eg^2}\right),
\label{eq:app-sxy-lowfreq}\\
\sxx(\ii\xi)&=\frac{\xi}{4\pi}\,\chitwoD_a(0)+\mathcal{O}(\xi^3).
\label{eq:app-sxx-lowfreq}
\end{align}
Similarly, we expand each reflection coefficient up to order $\mathcal{O}(\sigma)$. The denominator
differs from unity at
\begin{equation}
\frac{1}{\mathcal R}
=1-2\pi\tilde\sigma_{xx}\!\left(\lambda+\frac1\lambda\right)+\mathcal O(\sigma^2),
\label{eq:app-denom-expand}
\end{equation}
so that, inserting \eqref{eq:app-denom-expand} into
\eqref{eq:app-TMxx}--\eqref{eq:app-TMxy} and discarding all terms of order
$\sigma^2$,
\begin{align}
r_{xx}&=-2\pi\,\frac{\tilde\sigma_{xx}}{\lambda}+\mathcal O(\sigma^2),
\label{eq:app-rxx-lin}\\
r_{yy}&=-2\pi\,\tilde\sigma_{xx}\,\lambda+\mathcal O(\sigma^2),
\label{eq:app-ryy-lin}\\
r_{xy}&=-r_{yx}=-2\pi\,\tilde\sigma_{xy}+\mathcal O(\sigma^2).
\label{eq:app-rxy-lin}
\end{align}
We adopt the Gaussian normalization $\tilde\sigma=\sigma/c$ of
Ref.~\cite{TseMacDonald2012} throughout. 

The Casimir energy Eq.~\eqref{eq:lifshitz} depends on two scattering matrices of bodies $a=1$ and $a=2$. One reflection matrix is calculated inverting the sign of the scattering direction, a requirement denoted by a prime ($R'_a$) in Eq.~\eqref{eq:lifshitz}.
The Casimir energy is invariant under a basis change of the reflection matrices. However, the form of the matrices can change. While in the linear polarization basis, $(x,y)$, one can show $R'_a =R_a$, this is no longer true in the TE/TM basis~\cite{FKV2018}. This difference led an earlier calculation of the Casimir effect to mistakenly conclude that the force is repulsive between two Chern insulators  with opposite Chern numbers~\cite{Rodriguez-Lopez2014}, conflicting with e.g. Refs.~\cite{TseMacDonald2012,FKV2018}. As pointed out in Ref.~\cite{FKV2018} (Sec.~III\,B), Ref.~\cite{Rodriguez-Lopez2014} used
the Lifshitz kernel with $R$ in place of $R'$ in the TE/TM basis. Here we correct this issue and use the linear polarization basis. Our result below is fully consistent with Refs.~\cite{TseMacDonald2012,FKV2018} and predicts that two Chern insulator plates repel when they have equal Chern number.

\subsection{Expansion of the Lifshitz logarithm}
\label{app:logexpand}

To expand the {kernel} of Eq.~\eqref{eq:lifshitz} we write the argument of the determinant as $\mathbbm 1-M$ with
\begin{equation}
M\equiv R_1 R_2\,e^{-2\kappa d},
\label{eq:app-Mdef}
\end{equation}
and $\kappa=\sqrt{q^2+\xi^2/c^2}$.
Using $\ln\det X=\mathrm{Tr}\ln X$ and the series
$\ln(\mathbbm{1}-M)=-\sum_{k\ge1}M^k/k$,
\begin{equation}
\ln\det(\mathbbm 1-M)=-\,\mathrm{Tr}\,M-\half\,\mathrm{Tr}\,M^2-\cdots .
\label{eq:app-logdet}
\end{equation}
From the previous section, each reflection matrix is linear in the sheet conductivity,
$R_a=\mathcal O(\tilde\sigma^{(a)})$, so $\mathrm{Tr}\,M^k=\mathcal O[(\tilde\sigma^{(1)}
\tilde\sigma^{(2)})^k]\,e^{-2k\kappa d}$. 
The two sheet conductivities are expressed in terms of two small
parameters: $2\pi\tilde\sigma_{xy}=\alpha C=\mathcal O(\alpha)$ and  $2\pi\tilde\sigma_{xx}\sim\chitwoD/d\ll1$. The latter is small at the frequencies
$\xi\sim c/d$ that dominate the integral, since the kernel $e^{-2\kappa d}$ caps
$\kappa\lesssim1/d$. Hence the $k=1$ term dominates,
and
\begin{equation}
\ln\det(\mathbbm 1-M)\simeq-\mathrm{Tr}\,M=-\mathrm{Tr}[R_1R_2]\,e^{-2\kappa d}.
\label{eq:app-leading-log}
\end{equation}
Inserting \eqref{eq:app-leading-log} into Eq.~\eqref{eq:lifshitz} gives the leading contribution, Eq.~\eqref{eq:E2} of the main text,
\begin{equation}
\frac{E}{A}=-\frac{\hbar}{2\pi}\int_0^{\infty}\!\rd\xi
\int\!\frac{\rd^2 q}{(2\pi)^2}\,e^{-2\kappa d}\,\mathrm{Tr}[R_1R_2]+\cdots,
\label{eq:app-E2}
\end{equation}
which is bilinear in the sheet responses. 

\subsection{Evaluating the trace}

The trace can be written as
\begin{equation}
\mathrm{Tr}[R_1 R_2]=r^{(1)}_{xx}r^{(2)}_{xx}+r^{(1)}_{yy}r^{(2)}_{yy}
+r^{(1)}_{xy}r^{(2)}_{yx}+r^{(1)}_{yx}r^{(2)}_{xy}.
\label{eq:app-trace-general}
\end{equation}
The diagonal products are
\begin{equation}
r^{(1)}_{xx}r^{(2)}_{xx}
=4\pi^2\,\tilde\sigma^{(1)}_{xx}\tilde\sigma^{(2)}_{xx}\,\frac{\xi^2}{c^2\kappa^2},
\qquad
r^{(1)}_{yy}r^{(2)}_{yy}
=4\pi^2\,\tilde\sigma^{(1)}_{xx}\tilde\sigma^{(2)}_{xx}\,\frac{c^2\kappa^2}{\xi^2}.
\label{eq:app-diag-prod}
\end{equation}
Inserting the insulating form $\tilde\sigma^{(a)}_{xx}=\dfrac{\xi}{4\pi c}\,\chi_a$ with
$\chi_a\equiv\chitwoD_a(0)$, see main-text Eq.~\eqref{eq:sxx-lowfreq}, gives
%
\begin{equation}
r^{(1)}_{xx}r^{(2)}_{xx}+r^{(1)}_{yy}r^{(2)}_{yy}
=\tfrac14\,\frac{\xi^2}{c^2}\,\chi_1\chi_2
\left(\frac{\xi^2}{c^2\kappa^2}+\frac{c^2\kappa^2}{\xi^2}\right).
\label{eq:app-long-sum}
\end{equation}
Introducing the change of variables
\begin{equation}
x\equiv\frac{\xi}{c\kappa},\qquad 0\le x\le1,
\label{eq:app-xdef}
\end{equation}
so that $\xi^2/c^2=x^2\kappa^2$ we arrive at
\begin{equation}
r^{(1)}_{xx}r^{(2)}_{xx}+r^{(1)}_{yy}r^{(2)}_{yy}
=\tfrac14\,\chi_1\chi_2\,\kappa^2 x^2\!\left(x^2+\frac1{x^2}\right)
=\tfrac14\,\chi_1\chi_2\,\kappa^2\big(1+x^4\big).
\label{eq:app-long-final}
\end{equation}
\\
The off-diagonal products are calculated in a similar fashion. Using $r_{yx}=-r_{xy}$,
\begin{equation}
r^{(1)}_{xy}r^{(2)}_{yx}+r^{(1)}_{yx}r^{(2)}_{xy}
=-2\,r^{(1)}_{xy}r^{(2)}_{xy}
=-2\left(2\pi\tilde\sigma^{(1)}_{xy}\right)\!\left(2\pi\tilde\sigma^{(2)}_{xy}\right)
=-8\pi^2\,\tilde\sigma^{(1)}_{xy}\tilde\sigma^{(2)}_{xy}.
\label{eq:app-hall-step}
\end{equation}
Inserting the quantized value $\tilde\sigma^{(a)}_{xy}=C_a\alpha/(2\pi)$ 
\begin{equation}
r^{(1)}_{xy}r^{(2)}_{yx}+r^{(1)}_{yx}r^{(2)}_{xy}
=-8\pi^2\,\frac{C_1\alpha}{2\pi}\,\frac{C_2\alpha}{2\pi}=-2\alpha^2 C_1 C_2 .
\label{eq:app-hall-final}
\end{equation}
Combining \eqref{eq:app-long-final} and \eqref{eq:app-hall-final}, we obtain the total trace
\begin{equation}
\mathrm{Tr}[R_1 R_2]
=\tfrac14\,\chi_1\chi_2\,\kappa^2\big(1+x^4\big)-2\alpha^2 C_1 C_2 .
\label{eq:app-trace-final}
\end{equation}

\subsection{Phase-space integral}
\label{app:phasespace}

We change variables from $(\xi,q)$ to $(\kappa,x)$. From
$\kappa=\sqrt{q^2+\xi^2/c^2}$ and \eqref{eq:app-xdef},
\begin{equation}
\xi=c\kappa x,\qquad q=\kappa\sqrt{1-x^2}.
\label{eq:app-xi-q}
\end{equation}
The momentum integral is isotropic,
$\int\rd^2 q/(2\pi)^2=\int_0^\infty q\,\rd q/(2\pi)$. The Jacobian of the map is
\begin{equation}
\frac{\partial(\xi,q)}{\partial(\kappa,x)}
=\det\!\begin{pmatrix}
cx & c\kappa\\[2pt]
\sqrt{1-x^2} & -\dfrac{\kappa x}{\sqrt{1-x^2}}
\end{pmatrix}
=-\frac{c\kappa}{\sqrt{1-x^2}},
\label{eq:app-jacobian}
\end{equation}
so that, with $q=\kappa\sqrt{1-x^2}$,
$q\,\rd q\,\rd\xi=|\partial(\xi,q)/\partial(\kappa,x)|\,q\,\rd\kappa\,\rd x
=c\kappa^2\,\rd\kappa\,\rd x$, and
\begin{equation}
\frac{\rd\xi\,\rd^2 q}{(2\pi)^2}=\frac{c\,\kappa^2\,\rd\kappa\,\rd x}{2\pi},
\qquad \kappa\in[0,\infty),\ x\in[0,1].
\label{eq:app-measure}
\end{equation}
Substituting \eqref{eq:app-trace-final} and \eqref{eq:app-measure} into
\eqref{eq:app-E2} separates the energy into the two terms of Eq.~\eqref{eq:E-combined}. The factor
$e^{-2\kappa d}$ is independent of $x$.

\paragraph*{Dielectric term.}
The longitudinal part of \eqref{eq:app-trace-final} gives
\begin{align}
\frac{E_{\mathrm{diel}}}{A}
&=-\frac{\hbar}{2\pi}\int_0^1\!\rd x\int_0^\infty\!\rd\kappa\,
\frac{c\kappa^2}{2\pi}\,e^{-2\kappa d}\,
\tfrac14\,\chi_1\chi_2\,\kappa^2(1+x^4)
\nonumber\\
&=-\frac{\hbar c}{16\pi^2}\,\chi_1\chi_2
\int_0^1\!\rd x\,(1+x^4)\int_0^\infty\!\rd\kappa\,\kappa^4 e^{-2\kappa d}.
\label{eq:app-Ediel-int}
\end{align}
The angular integral is $\int_0^1(1+x^4)\,\rd x=\tfrac65$, and the radial integral is
$\int_0^\infty\kappa^4 e^{-2\kappa d}\rd\kappa=4!/(2d)^5=3/(4d^5)$. Hence
\begin{equation}
\frac{E_{\mathrm{diel}}}{A}
=-\frac{\hbar c}{16\pi^2}\,\chi_1\chi_2\cdot\frac65\cdot\frac{3}{4d^5}
=-\frac{9}{160\pi^2}\,\frac{\hbar c}{d^5}\,\chitwoD_1(0)\,\chitwoD_2(0),
\label{eq:app-Ediel}
\end{equation}
the first (attractive) term of Eq.~\eqref{eq:E-combined}.

\paragraph*{Hall term.}
The Hall part of \eqref{eq:app-trace-final} gives
\begin{align}
\frac{E_{\mathrm{Hall}}}{A}
&=-\frac{\hbar}{2\pi}\int_0^1\!\rd x\int_0^\infty\!\rd\kappa\,
\frac{c\kappa^2}{2\pi}\,e^{-2\kappa d}\,(-2\alpha^2 C_1 C_2)
\nonumber\\
&=+\frac{\hbar c\,\alpha^2 C_1 C_2}{2\pi^2}
\int_0^1\!\rd x\int_0^\infty\!\rd\kappa\,\kappa^2 e^{-2\kappa d}.
\label{eq:app-EHall-int}
\end{align}
Here $\int_0^1\rd x=1$ and
$\int_0^\infty\kappa^2 e^{-2\kappa d}\rd\kappa=2!/(2d)^3=1/(4d^3)$, so
\begin{equation}
\frac{E_{\mathrm{Hall}}}{A}
=+\frac{\hbar c\,\alpha^2 C_1 C_2}{2\pi^2}\cdot\frac{1}{4d^3}
=+\frac{\hbar c\,\alpha^2}{8\pi^2 d^3}\,C_1 C_2 ,
\label{eq:app-EHall}
\end{equation}
the second term of Eq.~\eqref{eq:E-combined}, which is repulsive for $C_1C_2>0$. 
Adding \eqref{eq:app-Ediel} and \eqref{eq:app-EHall} reproduces Eq.~\eqref{eq:E-combined},
\begin{equation}
\frac{E}{A}=
-\frac{9}{160\pi^2}\,\frac{\hbar c}{d^5}\,\chi_1\chi_2
+\frac{\hbar c\,\alpha^2}{8\pi^2 d^3}\,C_1 C_2+\cdots .
\label{eq:app-combined}
\end{equation}
As a check, the Hall pressure
$P_{\mathrm{Hall}}=-\partial_d(E_{\mathrm{Hall}}/A)=3\hbar c\,\alpha^2 C_1C_2/
(8\pi^2 d^4)$ reproduces the quantized Casimir force
$3\hbar c\,\alpha^2/(8\pi^2 d^4)$ of Ref.~\cite{TseMacDonald2012}.

The dielectric term~\eqref{eq:app-Ediel} admits an analogous check against the
longitudinal contribution detailed in the supplement of Ref.~\cite{TseMacDonald2012}. There the result is
written in terms of the derivative of the longitudinal conductivity on the imaginary axis,
$\rd\tilde\sigma_{xx}/\rd\omega|_0$, rather than the static polarizability $\chi$. The
two encode the same low-frequency response and are related through the insulating
form~\eqref{eq:sxx-lowfreq}: since $\sigma_{xx}(\ii\xi)=(\xi/4\pi)\chi$ and
$\tilde\sigma_{xx}=\sigma_{xx}/c$, one has $\tilde\sigma_{xx}(\ii\xi)=\xi\,\chi/(4\pi c)$,
linear in $\xi$, so that
\begin{equation}
\left.\frac{\rd\tilde\sigma_{xx}}{\rd\omega}\right|_{0}=\frac{\chi}{4\pi c}
\quad\Longleftrightarrow\quad
\chi=4\pi c\left.\frac{\rd\tilde\sigma_{xx}}{\rd\omega}\right|_{0}.
\label{eq:app-chi-slope}
\end{equation}
Substituting $\chi_a=4\pi c\,(\rd\tilde\sigma^{(a)}_{xx}/\rd\omega)|_0$ into
\eqref{eq:app-Ediel},
\begin{equation}
\frac{E_{\mathrm{diel}}}{A}
=-\frac{9}{160\pi^2}\frac{\hbar c}{d^5}\,(4\pi c)^2
\left.\frac{\rd\tilde\sigma^{(1)}_{xx}}{\rd\omega}\right|_0
\left.\frac{\rd\tilde\sigma^{(2)}_{xx}}{\rd\omega}\right|_0
=-\frac{9}{10}\frac{\hbar c^3}{d^5}
\left.\frac{\rd\tilde\sigma^{(1)}_{xx}}{\rd\omega}\right|_0
\left.\frac{\rd\tilde\sigma^{(2)}_{xx}}{\rd\omega}\right|_0,
\label{eq:app-Ediel-TM}
\end{equation}
which is precisely the longitudinal term in the supplement of Ref.~\cite{TseMacDonald2012}. 



\section{Exact all-distance benchmark: two saturating Landau-level plates}
\label{app:LLbenchmark}

The Landau-level example of the main text saturates the susceptibility bound
exactly. In this appendix we show that for parabolic Landau levels the
Lifshitz energy can in fact be evaluated at all distances. The
resulting crossover distances quantifies the error incurred by the large-distance truncation used to derive the bounds of the main text.
For electrons in a Landau level, the optical conductivities in the imaginary frequency axis are given, in the units used in the main text, by
~\cite{Morimoto2009}
\begin{equation}
2\pi\tilde\sigma_{xx}(i\xi)=\alpha C\,\frac{u}{1+u^{2}},
\qquad
2\pi\tilde\sigma_{xy}(i\xi)=\alpha C\,\frac{1}{1+u^{2}},
\qquad u\equiv\frac{\xi}{\omega_c}.
\label{eq:sigmadimless}
\end{equation}
Equation \eqref{eq:sigmadimless} contains the two static properties used in the main
text: when $u\to0$, $\sigma_{xy}(0)=Ce^{2}/h$ and
$\chi^{2D}(0)=4\pi c\,\partial_\xi\tilde\sigma_{xx}|_{0}=\frac{2e^{2}|C|}{\hbar\omega_c}$.

We consider two identical sheets, $C_1=C_2=C$ and equal $\omega_c$, and work
in units $\hbar=c=\omega_c=1$, so lengths are measured in $\hbar c/E_g$ with
$E_g=\hbar\omega_c$. Because Eqs.~\eqref{eq:sigmadimless} are both proportional to the small parameter $\alpha$ we can make use of the same expansion of the Casimir energy as in Appendix \ref{app:logexpand}.
Repeating the steps but retaining the full
frequency dependence \eqref{eq:sigmadimless}, the linearized trace becomes, with $s\equiv2\pi\tilde\sigma_{xx}$,
$h\equiv2\pi\tilde\sigma_{xy}$ and $x=\xi/(c\kappa)$,
\begin{equation}
\mathrm{Tr}[R_1R_2]
= s^{2}\Big(x^{2}+\frac{1}{x^{2}}\Big)-2h^{2}
=\alpha^{2}C^{2}\,\frac{\kappa^{2}(1+x^{4})-2}{\big(1+\kappa^{2}x^{2}\big)^{2}},
\label{eq:traceLL}
\end{equation}
where we used $u=\kappa x$, so that $s^2 x^2=\alpha^2C^2\kappa^2x^4/(1+u^2)^2$
and $s^2/x^2=\alpha^2C^2\kappa^2/(1+u^2)^2$. Since
$s,h=\mathcal{O}(\alpha)$, linearizing the reflection coefficients
and truncating the log-determinant at first order is controlled by
$\alpha$ at all distances, up to a relative error of $\mathcal{O}(\alpha)$ and
$\mathcal{O}(\alpha^{2})$, respectively.
Inserting \eqref{eq:traceLL} into Eq.~\eqref{eq:app-E2} gives
\begin{equation}
\frac{E}{A}
=-\frac{\hbar c\,\alpha^{2}C^{2}}{4\pi^{2}}
\int_{0}^{\infty}\!d\kappa\,\kappa^{2}e^{-2\kappa d}\,f(\kappa),
\qquad
f(\kappa)\equiv\int_{0}^{1}\!dx\,
\frac{\kappa^{2}(1+x^{4})-2}{\big(1+\kappa^{2}x^{2}\big)^{2}}.
\label{eq:EofdLL}
\end{equation}
The dependence on $\alpha$, $C$ and $B$ has factored out completely: in units
of $\hbar c/E_g$ the energy profile, and in particular the location of any
sign change, is a {universal} function of $d$.

The $x$ integral in \eqref{eq:EofdLL} is analytical and results in 
\begin{equation}
f(\kappa)
=\frac{(\kappa^{2}+1)(\kappa^{2}-3)}{2\kappa^{3}}\,\arctan\kappa
\;+\;\frac{\kappa^{4}+3}{2\kappa^{2}(1+\kappa^{2})}.
\label{eq:fclosed}
\end{equation}

Inserting Eq.~\eqref{eq:fclosed} into \eqref{eq:EofdLL} the energy and pressure change sign where
\begin{equation}
\int_{0}^{\infty}\!d\kappa\,\kappa^{2}e^{-2\kappa {d^{\mathrm{ex}}_\times}}f(\kappa)=0,
\qquad
\int_{0}^{\infty}\!d\kappa\,\kappa^{3}e^{-2\kappa {d^{\mathrm{ex}}_F}}f(\kappa)=0,
\label{eq:crossoverdefs}
\end{equation}
whose roots can be found numerically
\begin{equation}
{d^{\mathrm{ex}}_\times}=1.02148\,\frac{\hbar c}{E_g},
\qquad
{d^{\mathrm{ex}}_F}=1.38309\,\frac{\hbar c}{E_g}.
\label{eq:exactcrossovers}
\end{equation}
These are to be compared with the values obtained analytically in the main text by truncating to  the two static inputs $\chi(0)$ and $\sigma_{xy}(0)$:
the energy zero at ${d_\times}=2\sqrt{9/20}\,\hbar c/E_g\simeq1.342\,\hbar c/E_g$
 and the force zero at
${d_F}=\sqrt{3}\,\hbar c/E_g\simeq1.732\,\hbar c/E_g$ [Eq.~\eqref{eq:LL-dcross} of the main text].
As noted in the main text, the exact crossovers lie
{below} their truncated counterparts by $\sim25\%$: at $d\sim\hbar c/E_g$
the true $\sigma_{xx}(i\xi)$ falls below the linear-slope extrapolation
$\xi\chi(0)/4\pi$, so the frequency truncation overestimates the attraction.

\section{Flat-band Chern-model benchmark for the Casimir onset}
\label{app:flatband-casimir-benchmark}

This appendix gives the numerical benchmark used in the main text to illustrate the
role of the longitudinal polarizability in the large-distance Casimir interaction.
The calculation uses the flat-band Chern model of Sec.~5.7.3 of
Passos and Souza~\cite{Passos2026}, namely the three-orbital square-lattice
model of Yang \emph{et al.}~\cite{YGS12}.  We set
\begin{equation}
  t_1=1,\qquad
  t_2=-\frac{\Delta}{\sqrt{3}},\qquad
  \phi=\frac{\pi}{3},
  \label{eq:flatband-parameters}
\end{equation}
and place the chemical potential in the lower gap, so only the lowest band is
occupied.  The occupied band has Chern number $C=3$ throughout the range shown. The parameter $\Delta$ sets the flatness of the lowest band, which is maximal close to $\Delta = 1$.

\subsection{Bloch Hamiltonian}

With orbital labels $l=1,2,3$ understood modulo three, the Bloch Hamiltonian is
specified by
\begin{align}
  F_{\mathbf{k}}(\Phi)
    &= 2t_2 \cos(k_x+k_y-\Phi),                                      \\
  G_{1,\mathbf{k}}(\Phi)
    &= t_1\left(e^{i k_x}+e^{-i k_y+i\Phi}\right),                    \\
  G_{2,\mathbf{k}}(\Phi)
    &= t_2 e^{i k_x-i k_y+i\Phi}.
  \label{eq:flatband-FG}
\end{align}
The nonzero matrix elements are
\begin{align}
  H_{ll}(\mathbf{k})
    &= F_{\mathbf{k}}\!\left[(2l-1)\phi\right],                       \\
  H_{l+1,l}(\mathbf{k})
    &= G_{1,\mathbf{k}}(2l\phi),                                      \\
  H_{l+2,l}(\mathbf{k})
    &= G_{2,\mathbf{k}}\!\left[(2l+1)\phi\right],
  \label{eq:flatband-H-elements}
\end{align}
together with Hermitian conjugates.  

\subsection{Susceptibility and Chern bound}

The solid line in Fig.~\ref{fig:flatband-appendix-benchmark}(a) is the static
longitudinal polarizability.  In the notation of Passos and Souza, it is obtained
from their Eq.~(3c) at $p=-1$, together with their optical sum rule Eq.~(10) and the
$p=-1$ row of their Table~1, which identifies the real symmetric moment with the
clamped-ion susceptibility.  In the conventions used here, $\hbar=e^2=1$, this gives
\begin{equation}
  \chi^{2D}_{\alpha\alpha}(0;\Delta)
  =
  2
  \int_{\mathrm{BZ}}\frac{d^2k}{(2\pi)^2}
  \sum_{m\in\mathrm{emp}}
  \frac{
  \left|
  \left\langle u_{m\mathbf{k}}\middle|
  \partial_{k_\alpha}H_{\mathbf{k}}
  \middle|u_{0\mathbf{k}}\right\rangle
  \right|^2
  }{
  \left(\varepsilon_{m\mathbf{k}}-\varepsilon_{0\mathbf{k}}\right)^3
  },
  \label{eq:flatband-kubo-chi}
\end{equation}
where the occupied band is $n=0$ and $m=1,2$ are empty.  The model is $C_4$
symmetric, and numerically $\chi^{2D}_{xx}=\chi^{2D}_{yy}$ to the accuracy of the
mesh.  We therefore plot
\begin{equation}
  \chi^{2D}(0;\Delta)
  =
  \frac{1}{2}
  \left[
  \chi^{2D}_{xx}(0;\Delta)+\chi^{2D}_{yy}(0;\Delta)
  \right].
  \label{eq:flatband-chi-average}
\end{equation}

The dashed line in Fig.~\ref{fig:flatband-appendix-benchmark}(a) is the Chern
susceptibility bound of Passos and Souza.  The starting point is their Eq.~(109),
specialized to two dimensions,
\begin{equation}
  \chi^{2D}(0)
  \ge
  \chi_{\min}^{2D}
  =
  \frac{C^2}{\pi a_\perp^* n_e}.
  \label{eq:flatband-PS-bound}
\end{equation}
For a tight-binding model, the Bohr radius is replaced by their Eq.~(112),
\begin{equation}
  a_\perp^*
  =
  \frac{4\pi\epsilon_0\hbar^2}{e^2 \langle m^*\rangle_\perp}.
  \label{eq:flatband-PS-bohr}
\end{equation}
Equivalently, using their optical inverse-mass definition, Eq.~(24), and the
oscillator-strength sum rule, Eq.~(B.4), the bound becomes in the present lattice
units
\begin{equation}
  \chi_{\min}^{2D}(\Delta)
  =
  \frac{C^2}{4\pi^2 n_e\langle m^{*-1}\rangle_\perp}.
  \label{eq:flatband-bound-internal}
\end{equation}
There is one occupied spinless state per unit cell, so $n_e=1$ for lattice constant
$a=1$.  The inverse optical mass entering Eq.~\eqref{eq:flatband-bound-internal} was
computed from the same transition weights as Eq.~\eqref{eq:flatband-kubo-chi},
\begin{equation}
  \langle m^{*-1}\rangle_{\alpha\alpha}
  =
  \int_{\mathrm{BZ}}\frac{d^2k}{(2\pi)^2}
  \sum_{m\in\mathrm{emp}}
  \frac{
  2
  \left|
  \left\langle u_{m\mathbf{k}}\middle|
  \partial_{k_\alpha}H_{\mathbf{k}}
  \middle|u_{0\mathbf{k}}\right\rangle
  \right|^2
  }{
  \varepsilon_{m\mathbf{k}}-\varepsilon_{0\mathbf{k}}
  }.
  \label{eq:flatband-optical-mass}
\end{equation}
The plotted quantity in panel (a) is $4\pi\chi^{2D}$ and $4\pi\chi_{\min}^{2D}$,
which matches the atomic-unit normalization of Fig.~9 of
Ref.~\cite{Passos2026}.  Directly integrating the transport mass
$\partial^2_{k_\alpha}\varepsilon_0$ over a filled tight-binding band gives zero by
Brillouin-zone periodicity; the finite bound in Eq.~\eqref{eq:flatband-bound-internal}
uses the optical inverse mass instead.

\subsection{Casimir force sign-change distance}

For two identical sheets, the leading large-distance energy per area is
\begin{equation}
  \frac{E}{A}
  =
  -\frac{9}{160\pi^2}\frac{\hbar c}{d^5}
  \left[\chi^{2D}(0;\Delta)\right]^2
  +
  \frac{\hbar c\alpha^2}{8\pi^2d^3}C^2 .
  \label{eq:flatband-large-distance-energy}
\end{equation}
The first term is the attractive longitudinal contribution, and the second term is
the repulsive Hall contribution for equal-sign Chern numbers.  Since
$P=-\partial(E/A)/\partial d$, the pressure changes sign at
\begin{equation}
  d_F(\Delta)
  =
  \frac{\sqrt{3}}{2}
  \frac{\chi^{2D}(0;\Delta)}{\alpha |C|}.
  \label{eq:flatband-dF}
\end{equation}
Replacing $\chi^{2D}$ by Eq.~\eqref{eq:flatband-bound-internal} gives the lower
envelope
\begin{equation}
  d_F^{\min}(\Delta)
  =
  \frac{\sqrt{3}}{2}
  \frac{\chi_{\min}^{2D}(\Delta)}{\alpha |C|}.
  \label{eq:flatband-dFmin}
\end{equation}

\begin{figure*}[t]
  \centering
  \includegraphics[width=0.95\textwidth]{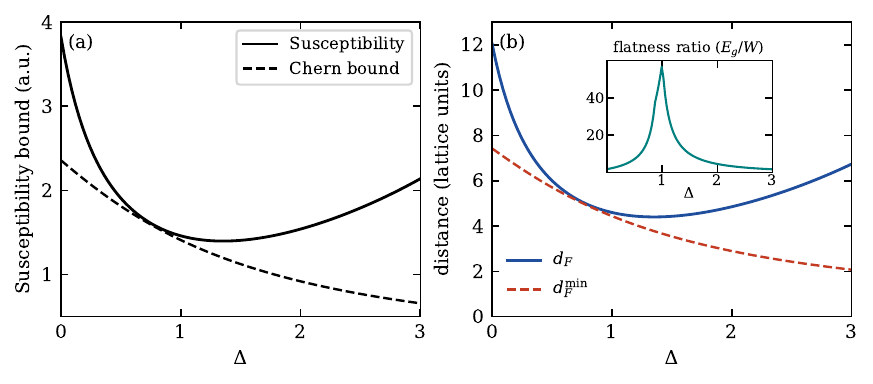}
  \caption{
  Flat-band Chern-model benchmark.  (a) Reproduction of the right panel of
  Fig.~9 of Passos and Souza~\cite{Passos2026} for the three-band $C=3$
  square-lattice model.  The solid line is the Kubo susceptibility
  Eq.~\eqref{eq:flatband-kubo-chi}; the dashed line is the Chern susceptibility
  bound, Eq.~\eqref{eq:flatband-bound-internal}, obtained from
  Eqs.~\eqref{eq:flatband-PS-bound} and \eqref{eq:flatband-PS-bohr}.  Both curves
  are multiplied by $4\pi$ to match the atomic-unit axis of Ref.~\cite{Passos2026}.
  (b) Force sign-change distance from Eq.~\eqref{eq:flatband-dF} and the lower
  envelope from Eq.~\eqref{eq:flatband-dFmin}.  The Hall contribution is fixed by
  $C=3$; the variation of $d_F$ with $\Delta$ is caused by the longitudinal
  polarizability. The inset shows the flatness ratio of the lowest band as a function of $\Delta$. Panel (b) is presented in the main text.}
  \label{fig:flatband-appendix-benchmark}
\end{figure*}

\subsection{Numerical details and interpretation}

The final scan used an $81\times81$ uniform Brillouin-zone mesh and 101 values of
$\Delta$ in the interval $0\le\Delta\le3$.  The computed Chern number remained
$C=3$ across the scan.
The flatness ratio is largest near $\Delta=0.99$, where
$\chi/\chi_{\min}=1.034$.  The closest approach to the bound occurs slightly below
the maximally flat point, at $\Delta=0.78$, where $\chi/\chi_{\min}=1.009$.  These
numbers reproduce the near saturation shown in Fig.~9 of Ref.~\cite{Passos2026}.

For the Casimir problem, the actual sign-change distance has a shallow minimum
$d_F=4.40$ at $\Delta=1.35$, while at the maximally flat point
$d_F\simeq4.61$.  By comparison, the wide-band point $\Delta=3$ gives
$d_F\simeq6.73$, and $\Delta=0$ gives $d_F\simeq12.0$, all in lattice-constant
units.  Thus the numerical benchmark shows that flattening lowers the onset distance by reducing
the longitudinal polarizability toward the Chern susceptibility bound, thereby
weakening the attractive contribution in Eq.~\eqref{eq:flatband-large-distance-energy}.

\section{Twisted semiconductor bilayers}
\label{app:tmote2}
 
Twisted semiconductor bilayers, most prominently twisted MoTe$_2$
(t-MoTe$_2$), realize quantized and fractionally quantized Hall
responses at zero magnetic field
[\onlinecite{Cai2023,Zeng2023,Park2023}]. Here we evaluate the
crossover bound \eqref{eq:dcross-identical} with realistic estimates for the charge density, effective mass and gap to quantify how far the system
is from saturating the susceptibility bound \eqref{eq:susc-bound}.

\emph{Density.} The carrier density is set by the moir\'e cell rather
than the atomic lattice. With the MoTe$_2$ lattice constant
$a_0 = 3.52\,$\AA{} and twist angle $\theta$, the moir\'e period is
$a_m = a_0/[2\sin(\theta/2)]$ and one hole per cell gives
$n_e = [(\sqrt{3}/2)\, a_m^2]^{-1}$. At $\theta = 3.7^{\circ}$,
$a_m \simeq 5.45\,$nm and
$n_e \simeq 2-4\times 10^{12}\,\mathrm{cm}^{-2}$
\cite{Cai2023,Park2023}. 
 
\emph{Optical mass.} The mass entering the renormalized Bohr radius
$a^{*}_{\perp} = \hbar^2/(e^2 m^*)$ is the optical mass of the
$f$-sum rule. In the continuum description of t-MoTe$_2$ the kinetic
term is $(\mathbf{k}-\mathbf{K}_{\pm})^2/2m^*$ with the monolayer
K-valley hole mass $m^* = 0.62\, m_e$ [\onlinecite{Wu2019}], while the
moir\'e potential and interlayer tunneling carry no momentum
dependence. The $f$-sum rule, and hence $a^{*}_{\perp}$, is therefore fixed
by the bare monolayer mass: the strong flattening of the
moir\'e band renormalizes the transport mass but not the optical one.
We use $m^* = 0.62\,m_e$, i.e.\ $a^{*}_{\perp} \simeq 0.085\,$nm.

\emph{Gap.} The charge gap of the $\nu=-1$ state,
measured locally by nanoSQUID magnetometry, is
$\Eg \simeq 14\,$meV \cite{Redekop2024}.\\

With these inputs, and for the mentioned electron density range $n_e \simeq 2-4\times 10^{12}\,\mathrm{cm}^{-2}$, the Chern susceptibility bound \eqref{eq:susc-bound}
gives
\begin{equation}
\chi_{\min} = \frac{C^2}{\pi a^{*}_{\perp} n_e} \simeq 192-96\,\mathrm{nm},
\label{eq:chimin-tmote2}
\end{equation}
and the crossover bound for two identical sheets,
Eq.~\eqref{eq:dcross-identical}, evaluates to
\begin{equation}
d_{\times} \geq \frac{3}{\sqrt{20}\,\pi}\,
\frac{|C|}{\alpha\, a^{*}_{\perp} n_e}
\simeq 17.6-8.8\,\mu\mathrm{m},
\qquad
d_{F} \geq \sqrt{\tfrac{5}{3}}\, d_{\times} \simeq 22.8-11.4\,\mu\mathrm{m}.
\label{eq:dcross-tmote2-app}
\end{equation}

This interval is of the order of $\hbar c/\Eg \simeq 14\,\mu$m which controls the large distance expansion. 
By the same reasoning as discussed in the main text, the
truncation overestimates the attraction, so
\eqref{eq:dcross-tmote2-app} falls on the conservative side. 

\end{widetext}

\end{document}